\documentclass[12pt]{article}
\pdfoutput=1
\usepackage{latexsym, graphicx,cite} 
\usepackage{amsmath}
\usepackage{amssymb}
\usepackage{tikz-feynman}
\allowdisplaybreaks
\usepackage[top = 1. in,bottom = 1. in, left = 1 in, right=1 in]{geometry}

\newcommand{\arXiv}[1]{\href{http://www.arXiv.org/abs/#1}{arXiv:#1}}
\usepackage[colorlinks=true, linkcolor=blue, bookmarks=true]{hyperref}

\makeatletter
\renewcommand\section{\@startsection {section}{1}{\z@}%
                                   {-3.5ex \@plus -1ex \@minus -.2ex}
                                   {2.3ex \@plus.2ex}%
                                   {\normalfont\large\bfseries}}
\renewcommand\subsection{\@startsection{subsection}{2}{\z@}%
                                     {-3.25ex\@plus -1ex \@minus -.2ex}%
                                     {1.5ex \@plus .2ex}%
                                     {\normalfont\bfseries}}
\makeatother
\newcommand{\beq}{\begin{equation}}
\newcommand{\eeq}{\end{equation}}
\newcommand{\ber}{\begin{array}}
\newcommand{\eer}{\end{array}}
\newcommand{\del}{\partial}

\newcommand{\ssty}{\scriptstyle}
\newcommand{\dsty}{\displaystyle}
\newcommand{\s}{\sigma}
\newcommand{\te}{\theta}

\newcommand{\de}{\delta}
\newcommand{\De}{\Delta}
\newcommand{\eps}{\epsilon}
\newcommand{\cnst}{\mbox{const}}
\newcommand{\om}{\omega}
\newcommand{\al}{\alpha}
\newcommand{\ab}{\bar\alpha}

\newcommand{\ad}{a^\dagger}

\begin{document}
\begin{titlepage}
\begin{center}
{\LARGE\bf Resonant Hamiltonian systems and weakly\vspace{2mm}\\nonlinear dynamics in AdS spacetimes}    \\
\vskip 8mm
{\large Oleg Evnin}
\vskip 3mm
{\em Department of Physics, Faculty of Science, Chulalongkorn University, Bangkok, Thailand}
\vskip 2mm
{\em Theoretische Natuurkunde, Vrije Universiteit Brussel and\\
The International Solvay Institutes, Brussels, Belgium}
\vskip 3mm
{\small\noindent  {\tt oleg.evnin@gmail.com}}
\vskip 8mm
\end{center}

\noindent {\bf Abstract:} Weakly nonlinear dynamics in anti-de Sitter (AdS) spacetimes is reviewed, keeping an eye on the AdS instability conjecture and focusing on the resonant approximation that accurately captures in a simplified form the long-term evolution of small initial data. Topics covered include turbulent and regular motion, dynamical recurrences analogous to the Fermi-Pasta-Ulam phenomena in oscillator chains, and relations between AdS dynamics and nonrelativistic nonlinear Schr\"odinger equations in harmonic potentials. Special mention is given to the way the classical dynamics of weakly nonlinear strongly resonant systems is illuminated by perturbative considerations within the corresponding quantum theories, in particular, in relation to quantum chaos theory.

\tableofcontents 

\vfill

\end{titlepage}

\section{Introduction}

This review grew as an expanded version of my presentation at the BIRS workshop ``Time-like Boundaries in General Relativistic Evolution Problems'' held in the Mexican city of Oaxaca in 2019. It is my pleasure to
start out by expressing my cordial gratitude to the workshop program at Casa Matem\'atica Oaxaca and the principal organizers Olivier Sarbach, Piotr Bizo\'n, Helmut Friedrich and Oscar Reula, as well as all of the participants, for putting together a lovely event that has been both stimulating intellectually and rewarding socially at a charming location suffused with history and tradition.

It deserves to be mentioned at the start how the topic of this review, which is most broadly weakly nonlinear dynamics
in strongly resonant bounded domains, and more specifically this type of dynamics in Anti-de Sitter (AdS) spacetimes, connects to the scope of the workshop. Time-like boundaries are indeed what allows a bounded
domain to form, eliminates scattering to infinity, and allows nonlinearities, no matter how small, to affect the dynamics over an unlimited duration of time.
Perhaps the simplest general-relativistic example is the gravitational dynamics of space inside a spherical cavity considered in \cite{cavity}.

A very prominent realization of this cavity-type dynamics in a general-relativistic context is given by the global Anti-de Sitter spacetime, which is remarkable in a number of ways.
First, it is a maximally symmetric spacetime that both stands out mathematically and provides a solution to vacuum Einstein's equations
with a negative cosmological constant. Second, it has been playing a central role in the subject of AdS/CFT correspondence, a significant
line of research in theoretical high-energy physics of the recent decades (a review can be found in \cite{AdSrev}). Third, it possesses
an extremely resonant spectrum of linearized mode frequencies, which is dictated by the structure of its isometry group and
will play a crucial role in what follows. `Extremely resonant' in this context refers to the property that the diference of frequencies of any two normal modes for any free field in AdS is integer in appropriate units, which results in an infinite number of resonant relations between the frequencies, and strongly enhances the energy transfer between the normal modes in the weakly nonlinear regime.

AdS spacetimes do not possess an ordinary time-like boundary, but they possess a conformal time-like boundary that operates for massless fields just like an ordinary time-like boundary, and makes AdS function as a cavity. Crucially, the presence of the conformal boundary eliminates dispersal by scattering to spatial infinity that underlies the stability
of Minkowski space \cite{ChK}.

Mathematical generalities of setting up gravitational dynamics in asymptotically AdS spacetimes have been discussed in \cite{Friedrich,IW,Anderson,HS}.
The surge of interest in weakly nonlinear dynamics in AdS seen in the last decade has been triggered by the numerical and analytic
observations of \cite{BR} suggesting that black holes form in the evolution of a gravitating scalar field in AdS in spherical symmetry starting
from some arbitrarily small initial data. This has led the authors of \cite{BR} to formulate the {\it AdS instability conjecture,} hitherto unproved,
that an open set of initial perturbation profiles exists in AdS so that no matter how much the amplitude of these profiles is scaled down, a black hole 
will still form in the subsequent gravitational evolution. We emphasize that the instability in question
is specific to the `reflective' Dirichlet boundary conditions for the massless fields in AdS (this formulation is common in the AdS/CFT correspondence studies), and
does not have to be present for other boundary conditions \cite{Friedrich2}. See for example, \cite{dissipative} for considerations of linear fields in AdS with dissipative
boundary conditions geared toward a proof of AdS stability with such boundary conditions, and \cite{robin} for stability studies with Robin boundary conditions.

One word that comes essentially into the considerations of AdS instability is `turbulence.' The term, as employed here, means transfer of energy
toward short-wavelength modes driven by nonlinearities in a purely deterministic setting, akin to similar studies of this type of dynamics for nonlinear Schr\"odinger equations that have been common in recent years in the mathematical PDE community, in particular, following the publication of \cite{nls}.
Such deterministic `weak turbulence' driven purely by nonlinearities is rather similar to `wave turbulence' \cite{nazarenko}, but without any statistical phase space averaging; the turbulent behavior is observed in a given, fully deterministic dynamical trajectory. It is more distantly related to the more familiar hydrodynamic turbulence, which is usually triggered by linear instabilities, rather than being purely due to nonlinear effects. All of these notions of turbulence, however, crucially involve transfer of energy towards shorter wavelengths (as one can witness with one's own eyes as small turbulent ripples form in initially smooth flows of a liquid, or streams of smoke).

In what manner does the notion of `weak turbulence' enter the discussion of AdS instability?
If a black hole is to form starting from a very small initial perturbation in AdS, heuristically, the energy of this perturbation must concentrate
in a very small region. This is only possible if short wavelength modes get appreciably excited -- arbitrarily short wavelenths, in fact,
if arbitrarily small black holes are to form starting from arbitrarily small initial perturbations. The transfer of energy from the initial fixed profile
with a given wavelength spectrum towards arbitrarily short wavelengths is precisely the turbulent cascade that must precede black hole formation.
Such processes have indeed been observed starting from \cite{BR}. Of course, the cascade is a necessary rather than sufficient condition for
black hole formation, but it is natural to approach the problem by studying it first.

An essential progenitor of the sophisticated weakly nonlinear dynamics in AdS is the highly resonant spectrum of linearized normal mode frequencies, observed for all conventional linear field systems
in AdS and originating from the structure of the AdS$_{d+1}$ isometry group $SO(d,2)$, where $d$ is the number of spatial dimensions. For fields of arbitrary mass and spin in AdS, differences of frequencies of any
two linearized normal modes are integer in appropriate units (fixed in terms of the AdS curvature radius).
Without such resonances, one would expect that the effect of small nonlinearities should remain uniformly
small for all times in the spirit of the KAM theorem \cite{arnold} (though rigorously proving analogs of the
KAM theorem for systems with an infinite number of degrees of freedom may be challenging). By contrast,
in the presence of the enormous number of resonances characteristic of AdS, an enhancement of interactions occurs, so that arbitrarily small nonlinearities may induce effects of order 1 provided that one waits long enough. A typical situation one should have in mind here is considering the dynamics on time scales $1/g$ for a small nonlinear coupling $g$ (which is the same as considering time scales $1/\eps^2$ 
for initial perturbations of small amplitude $\eps$, as in \cite{BR}, when the leading nonlinearities in the equations of motion are cubic and $\eps$-independent). This is, incidentally, the same time scale on which black hole formation
was observed in the numerical simulations of \cite{BR}.

We thus set out to study the effects of nonlinearities in small solutions of amplitude $\eps$ in AdS
on time scales $1/\eps^2$. In this regime, an essential simplification can be employed in the equations
that we shall refer to as the {\it resonant approximation} (the same technique may be known as multiscale analysis or time-averaging \cite{murdock}, effective equation \cite{KM}, two-time framework \cite{FPU}
or renormalization flow equation \cite{CEV1,CEV2} depending on which community uses it; systematic textbook discussions can be found in \cite{murdock,KM}). In this approach, one first rewrites the nonlinear dynamical equations in the mode space using the linearized normal mode basis, and then observes
that these equations contain many rapidly oscillating terms whose cumulative effect can be bounded
and is suppressed by the effective nonlinearity magnitude, which gives grounds to discard them.
The remaining terms are precisely the ones whose effect may accumulate to large contributions on long time-scales, and these terms precisely correspond to interactions between modes whose frequencies satisfy rational resonance relations. On account of the highly resonant AdS spectrum, and in contrast to more conventional non-resonant cavities, the resonant approximation in AdS possesses a very rich dynamics
that includes strong energy transfer between the modes and turbulent phenomena. These dynamical solutions within the resonant approximation capture the weak field dynamics of AdS field systems,
and in particular the turbulent cascades leading to black hole formation that are the essence of AdS instability \cite{BR}.

The resonant approximation was introduced to studies of AdS instability problems in \cite{FPU}, primarily
as an alternative approach to numerical work, and then given a fully analytic formulation in \cite{CEV1,CEV2} (deriving a fully analytic form of gravitational AdS resonant systems is challenging on
account of the complexity of gravitational nonlinearities). Our purpose here will be to review the work
on resonant approximations to the AdS dynamics along the lines of \cite{FPU,FPUcomm,FPUrepl,CEV1,CEV2,dual,BMR,GMLL,UV1,UV2,UV3,Deppe,returns,AOQ,squashed}. The complexity of the resonant dynamics in AdS invites, furthermore, studies of simpler related systems, for example interacting scalars in nondynamical AdS
backgrounds \cite{BKS,CF,BHP,BEL}. Such systems both serve as `toy models' for the gravitational dynamics in AdS, and display rich and interesting dynamical phenomena in their own right, though turbulence appears less common in this setting than in the cases with gravitational interactions. One can furthermore take a nonrelativistic limit of AdS systems \cite{BEL,BEF}, obtaining nonlinear Schr\"odinger equations in harmonic potentials \cite{BEF,BMP,BBCE,nonrel1,nonrel2}. These equations possess resonant dynamics
very similar to AdS, and their resonant approximations have an identical algebraic structure; they are furthermore common in studies of trapped ultracold atomic gases \cite{BDZ}, far afield from the gravitational problems that brought us in contact with this range of mathematical problems.
There are also interesting dynamical systems designed {\it ad hoc} in the mathematical literature
\cite{GG,Xu,cascade} that look structurally identical to resonant systems originating from the weakly nonlinear AdS dynamics, except that the values of the mode couplings are different. These systems display various forms of integrability and some remarkable behaviors, and it is natural to review them together with the resonant systems originating from AdS, in search of a broader, more robust perspective.

We therefore aim to review the applications of resonant systems to studies of weakly nonlinear dynamics of bosonic fields in AdS, as well as other resonant systems emerging from various branches of physics and mathematics that share the mathematical structure of the AdS problems. 
To give the reader an impression of the type of dynamics that will be considered, a representative equation of motion is
\beq
i\,\frac{d\al_n}{d\tau}=\hspace{-3mm}\sum_{n+m=k+l} \hspace{-3mm}C_{nmkl}\,\ab_m\al_k\al_l,
\label{reseq}
\eeq
where $\al_n$ with $n\ge 0$ are complex dynamical variables that physically originate from complex amplitudes of linearized normal modes. Here and everywhere else in this review, bars (as in $\ab_n$) will denote complex conjugation.
The interaction coefficients $C$ encode the specific nonlinearities of the physical equations that the resonant system approximates,
while the {\it resonance condition} $n+m=k+l$ picks out only interactions between those quartets of modes whose frequencies are in resonance, as shall be seen in our explicit derivations that follow. These equations of motion can be derived from the Hamiltonian
\beq
H=\frac12\sum_{n+m=k+l} C_{nmkl}\,\ab_n\ab_m\al_k\al_l
\label{Hres}
\eeq
assuming that the conjugate momentum of $\al_n$ is $i\ab_n$, so that $\del H/\del(i\ab_n)$ gives $\del\al_n/\del\tau$ in accordance with (\ref{reseq}), and provided that $C_{nmkl}$ enjoy the index permutation symmetries $C_{nmkl}=C_{mnkl}=C_{nmlk}=\bar C_{klnm}$, the last equality required to ensure that the Hamiltonian is real-valued. These symmetries make $C$ a rank 4 {\it Hermitian tensor} in the mode space (such a notion of Hermitian tensors was introduced independently, and guided by different motivations, in \cite{htensor1,htensor2}).

The focus of the review will be on displaying the logical structure of the subject with a bare minimum of technical details, and giving references to the original literature from which the technical details may be recovered. Some mention will also be given to a number of heuristic ideas that have circulated in the community working on AdS dynamics over the years, without having been voiced in print. While these ideas are far from having been given a solid and precise form, it is useful to have them explicitly stated to stimulate future work on the subject. 

To keep the scope manageable, we focus
on dynamics within the resonant approximation only, to the exclusion of a large number of significant related research directions, in particular:
\begin{enumerate}
\item Direct numerical simulations of Einstein's equations with a negative cosmological constant, without applying the resonant approximation
\cite{JRB,BLL,BJ,AdSLMS,DKFK,revival,JGCh,JG,BFKR,CDSW,BR2,CDF,CMW,BFR}.
\item Construction of time-periodic solutions in AdS \cite{period,FFG,sks} that has been approached both by perturbative methods (where one fine-tunes the initial conditions to eliminate the secular terms at higher and higher orders of the perturbative expansion) and numerically. (We will, however, briefly touch on related considerations \cite{nonspher1a,nonspher1b,nonspher2,nonspher3,nonspher4,nonspher5,nonspher6,nonspher7,highpert} within the topic of gravitational dynamics in AdS outside spherical symmetry.)
\item Studies of the Einstein-Vlasov system in AdS \cite{Moschidis,AY} where powerful
instability results have been obtained in \cite{Moschidis} at a mathematically rigorous level. We note, however, that the dynamical realization of this instability is rather different from the classical field systems treated in \cite{BR} and the extensive follow-up work, nor is there a direct contact with the conventional approaches to the AdS/CFT correspondence.
\item `Stroboscopic' studies of the AdS dynamics \cite{DFLY}, where the evolution is viewed as a succession of linear bounces out to the AdS boundary and back (it is assumed that, as the wavefront expands, its amplitude goes down and nonlinearities effectively switch off) and nonlinear scattering during the epochs when the wavefront re-converges to the origin (where the AdS curvature is treated as negligible). This approach forms an intriguing alternative to the studies within the resonant approximation, and bears some similarities to the considerations of \cite{Moschidis}, but it is difficult to make mathematically precise. Similar considerations exist in the mathematical literature for nonlinear Schr\"odinger equations in harmonic potentials \cite{Carles}. 
\item Studies of AdS dynamics with periodic driving introduced via the boundary conditions at the conformal boundary \cite{floquet1,floquet2}. In \cite{cownden}, this problem was recast in a language closely reminiscent of the resonant approximations for AdS systems with `reflective' Dirichlet boundary conditions (the `zero driving' case that we consider here).
\item Studies of turbulence in relativistic conformal fluids \cite{fluid1,fluid2,fluid3}, connected to gravitational dynamics in AdS via the fluid/gravity correspondence \cite{AdSfluid}, a mathematical relation between Einstein's equations in asymptotically AdS spacetimes and Navier-Stokes equations for relativistic conformal fluids.
\end{enumerate}

The review is organized as follows: First, in section~\ref{sec2}, we revisit the derivation of the resonant
approximation to a cubic Klein-Gordon equation in AdS, which is a simple prototype for the treatment of gravitationally interacting fields. We then state, in section~\ref{sec3}, how this treatment can be adapted
to gravitational interactions in spherical symmetry. In spherical symmetry, gravity has no propagating degrees of freedom and the metric can be expressed through the matter field configuration on each time slice, producing a complicated nonlinear wave equation and naturally yielding to the treatment of section~\ref{sec2}. We briefly comment in section~\ref{sec3bis} on the situation outside spherical symmetry, where analytic considerations are very challenging and only a limited amount of progress has been made. Section~\ref{sec4} deals with the selection rules that make some mode coupling coefficients vanish \cite{CEV1,Yang,EN}, which is crucial for maintaining the structure of the resonant approximation and has far-reaching dynamical consequences \cite{CEV2,dual}, in particular, for massless fields.
The next three sections review three types of dynamics resonant Hamiltonian systems may manifest.
In section~\ref{sec5}, the focus is on turbulent cascades, which are at the heart of AdS instability,
both for the gravitationally interacting fields in AdS and for other related resonant systems.
Section~\ref{sec6} focuses on situations where some special solutions of the resonant system can be obtained and their mode amplitude spectrum is exactly periodic in time, indicating periodic energy transfer.
Such resonant systems arise from nonlinear wave equations in AdS, and may also be expected to emerge for gravitating fields, but only outside spherical symmetry. The general mathematical structure responsible for these behaviors has been laid out in \cite{solvable,breathing}. Section~\ref{sec7} deals with situations when periodicities in the mode amplitude evolution are no longer perfect, but very accurate returns to close proximity of the initial configurations occur \cite{FPU,returns} in the manner reminiscent of the Fermi-Pasta-Ulam recurrence phenomena in nonlinear oscillator chains \cite{FPUrev}. In section~\ref{sec8}, we review nonrelativistic systems that display close mathematical parallels to the AdS dynamics, while the main physical motivation for their studies comes from trapped ultracold atomic gases \cite{BDZ}. In section~\ref{sec9}, we take a step aside from the classical field-theoretic considerations that define our narrative and quantize resonant systems. This off-beat step results in Hamiltonian quantum systems of a striking simplicity \cite{quantres}, which are in fact so simple that they can be used as a heuristic tool to shed further light on the sophisticated classical resonant dynamics, in particular, in connection to the lore of the quantum chaos theory \cite{haake,DKPR}. We conclude with a summary and list some open problems.


\section{Weakly nonlinear Klein-Gordon equations in AdS}\label{sec2}

To illustrate the essentials of weakly nonlinear resonant dynamics in global AdS spacetimes, it is wise to turn to nonlinear Klein-Gordon equations for scalar fields,
which are relatively simple but thoroughly capture the relevant structures. In fact, the analysis of gravitational dynamics in spherical symmetry, to be reviewed in the next section, follows an identical pattern with more complicated nonlinearities, and even outside spherical symmetry, the example of nonlinear Klein-Gordon equations provides some guidance. Note that even in this simplified setting, the derivations are still relatively complicated. Readers who prefer to get a quick overview of how resonant Hamiltonian systems arise as approximations to weakly nonlinear strongly resonant dynamics of PDEs may choose to first consult the end of section~\ref{sec8} where analogous derivations, only a few lines long, are given for a much simpler (and closely related) nonrelativistic nonlinear Schr\"odinger equation.

The (global) AdS$_{d+1}$ spacetime with $d$ spatial dimensions and radius 1 can be realized as a hyperbolic hypersurface in a flat pseudo-Euclidean space of dimension $d+2$ with the coordinates $X^I=(X,Y,Z^1,\ldots,Z^d)$ and the metric $\eta_{IJ}=\mbox{diag}(-1,-1,1,\ldots,1)$, defined by the equation
\beq
\eta_{IJ}X^IX^J\equiv -X^2-Y^2+Z^iZ^i=-1.
\label{embedding}
\eeq
The homogeneous linear coordinate transformations that leave $\eta_{IJ}$ invariant by definition form the group $SO(d,2)$, and they also respect
the embedding equation (\ref{embedding}), thus forming an isometry group of AdS$_{d+1}$. The group $SO(d,2)$ also happens to be the conformal group of the $d$-dimensional Minkowski space, a relation that underlies the AdS/CFT correspondence \cite{AdSrev}. At a more basic level, counting the number of independent isometries reveals that it is maximal for the given dimension. AdS is thus a maximally symmetric spacetime, its curvature is constant, and it satisfies vacuum Einstein's equations with a negative cosmological constant $\Lambda$,
\beq
R_{\mu\nu}-\frac12 Rg_{\mu\nu}+\Lambda g_{\mu\nu}=0.
\eeq
One may parametrize the hyperboloid (\ref{embedding}) as
\beq
X=\frac{\cos t}{\cos x},\qquad Y=\frac{\sin t}{\cos x},\qquad Z^i=n^i(\Omega)\tan x,
\label{embed}
\eeq
where $x$ ranges from 0 to $\pi/2$, $n^i(\Omega)$ is the unit vector pointing in the direction given the usual hyperspherical angles on a $(d-1)$-sphere collectively denoted $\Omega\equiv\left\lbrace \theta_{1},\cdots, \theta_{d-2}, \varphi \right\rbrace$. The induced metric is then
\beq
ds^2=\frac{1}{\cos^2{x}}\left(-dt^2+dx^2+\sin^2{x}\,d\Omega_{d-1}^2\right),
\label{adsmetric}
\eeq
where $d\Omega_{d-1}^2$ is the line element of the round metric on a $(d-1)$-dimensional unit sphere parametrized by $\Omega$. As formulated, $t$ runs from 0 to $2\pi$, but the metric is $t$-independent, so the range of $t$ can be straightforwardly extended so that it runs from $-\infty$ to $\infty$, with the metric unchanged, and hence still in accord with vacuum Einstein's equations with a negative cosmological constant. The above metric makes it evident that AdS$_{d+1}$ is conformal to one half of the Einstein static universe $R\times S^d$ (cut by the equator). It follows that light rays can reach the conformal boundary $x=\pi/2$ in a finite time, which is one way to see that extra boundary conditions at $x=\pi/2$ must be prescribed in order to make the evolution well-defined.

We shall examine a cubic Klein-Gordon equation for a real scalar field $\phi$,
\beq
\Box\phi-m^2\phi=\phi^3,
\eeq
in the background (\ref{adsmetric}), where it becomes
\beq\label{AdSwave}
\cos^2{x}\left(-\del_t^2\phi+\frac{1}{\tan^{d-1}{x}}\del_x\left(\tan^{d-1}{x}\del_x\phi\right)+\frac{1}{\sin^2{x}}\De_{S^{d-1}}\phi\right)-m^2\phi=\phi^3,
\eeq
with $\De_{S^{d-1}}$ being the standard $(d-1)$-sphere Laplacian. We shall impose, here and elsewhere, the reflective Dirichlet boundary conditions
\beq
\phi\left(x\!=\!\frac{\pi}2\right)\,=\,0.
\eeq
These boundary conditions are simple and natural, and even more importantly, they are of central importance (together with their inhomogeneous modification where the right-hand side is replaced with a prescribed function) in the context of AdS/CFT correspondence. With these boundary conditions, a resonant spectrum of normal modes will emerge, opening the gates for a wealth of sophisticated weakly nonlinear phenomena.

To understand the weakly nonlinear dynamics of (\ref{AdSwave}) one must first study the linearized problem obtained by replacing the right-hand side with zero. This linearized problem can be solved by
separation of variables. One first introduces the mode functions that satisfy
\beq \label{eq_eigenvalAdS}
\left(\frac{1}{\tan^{d-1}{x}}\del_x\left(\tan^{d-1}{x}\del_x\right)+\frac{1}{\sin^2{x}}\De_{S^{d-1}}-\frac{m^2}{\cos^2{x}}\right)e_{nlk}(x,\Omega)=-\om_{nl}^2e_{nlk}(x,\Omega).
\eeq
As usual for systems with spherical symmetry, we assume that each $e_{nlk}$ is given by a function of $x$
times a hyperspherical harmonic $Y_{lk}$ of angular momentum $l$ satisfying
\beq
\De_{S^{d-1}}Y_{lk}=-l(l+d-2)Y_{lk}.
\eeq
Different angular momentum states within the same $l$-multiplet are labelled by the index $k$ ($k$ is often realized as a multi-index depending on which specific scheme one chooses for numbering the states within angular momentum multiplets). For each particular angular momentum state, $n$ gives the radial overtone number. One can then explicitly solve (\ref{eq_eigenvalAdS}) to obtain
\beq \label{eq_modefunctionAdS}
e_{nlk}(x,\Omega)=\mathcal{N}_{nlk}\cos^{\de}{x}\sin^l{x}P_n^{\left(\de-\frac{d}{2}, l+\frac{d}{2}-1\right)}(-\cos{2x})Y_{lk}(\Omega)
\eeq
and
\beq \label{eq_eigenvalueAdS}
\om_{nl}=\de+2n+l,
\eeq
where $\de=\frac{d}{2}+\sqrt{\frac{d^2}{4}+m^2}$ may be known as the {\it conformal dimension} of the field $\phi$, and $\mathcal{N}_{nlk}$ is a normalization factor. The mode functions are expressed through the Jacobi polynomials $P_n^{(a,b)}(x)$ \cite{Jacobi}. 
The fact that orthogonal polynomials define the structure of the mode functions will later have many
repercussions as it imparts special algebraic properties to the nonlinear mode couplings. The frequencies 
(\ref{eq_eigenvalueAdS}) do not depend on $k$ on account of rotational invariance. The mode functions are orthogonal with respect to the scalar product $\int dx\,d\Omega\,\tan^{d-1}x \,\bar e_{nlk}\, e_{n'l'k'}$.

\begin{figure}[t]
\centering
\includegraphics[scale=0.8]{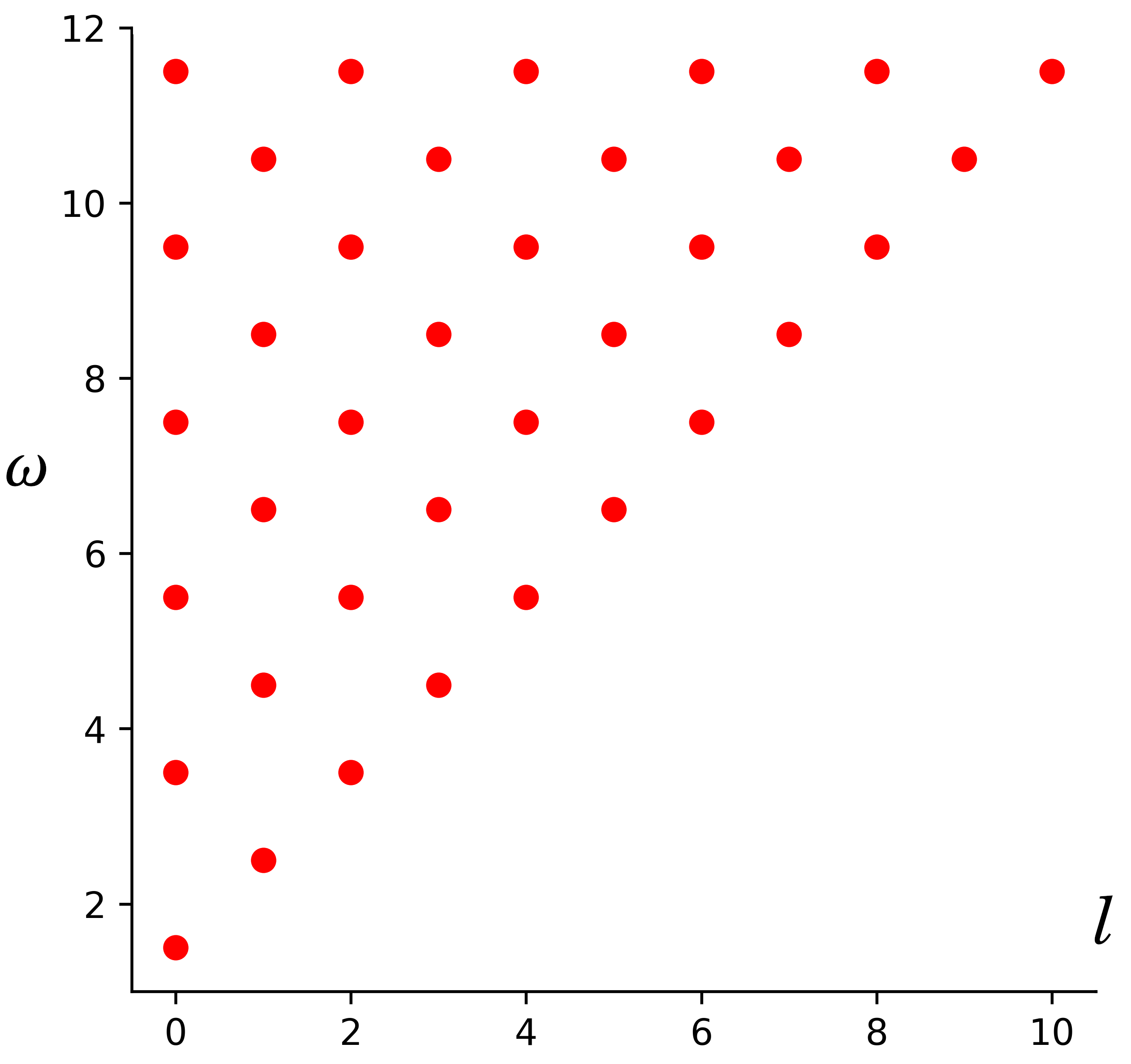}
\caption{The frequencies \eqref{eq_eigenvalueAdS} for different values of $n$ and $l$, and $\de=3/2$.}
\label{figtower}
\end{figure}

A crucial feature of the linearized normal modes is that their frequencies (\ref{eq_eigenvalueAdS}) form
a highly resonant spectrum. The difference of any two normal mode frequencies is integer for any given $\de$. This feature is in fact completely generic in AdS, and shared by fields of other spins. It can be traced back to the structure of the isometry group $SO(d,2)$.
In particular, there are many diophantine relations of the form
\beq
\sum_{nl} p_{nl} \om_{nl}=0
\label{resgen}
\eeq
with integer $p_{nl}$. Such relations are known to have a crucial impact on weakly nonlinear dynamics,
or more generally in the context of Hamiltonian perturbation theory \cite{arnold}.
We give a graphic representation of the tower of normal mode frequencies in fig.~\ref{figtower}.
Each dot in this figure represents a full angular momentum multiplet for the given value of $l$, with individual states, whose frequencies are degenerate, labelled by $k$.
One may notice that there is extra degeneracy beyond the angular momentum degeneracy dictated 
by the rotational symmetry (groups of dots are located along the same horizontal line).
This degeneracy may be seen as a hidden symmetry, which may be manifested by relating
 (\ref{eq_eigenvalAdS}) to the Schr\"odinger equation of a
superintegrable quantum-mechanical system \cite{hggs1,hggs2} known as the `Higgs oscillator' \cite{Higgs,Leemon}. (It is an interesting question how to generate other spacetimes with highly resonant frequency spectra. Such resonant domains are in general special and uncommon. For example, rectangular cavities with periodically identified boundaries are only weakly resonant, as in \cite{nls}, while the spherical cavity considered in \cite{cavity} is strongly resonant for spherically symmetric perturbations, but does not have similar properties outside spherical symmetry. While no general algorithm seems available, a recipe has been proposed in \cite{KG} for constructing such strongly resonant spacetimes out of a large class of superintegrable quantum-mechanical systems in a way that generalizes the relation of AdS and the Higgs oscillator.)
With these preliminaries, the general linearized solution of (\ref{AdSwave}) can be written as
\beq\label{AdSlin}
\phi_{\mbox{\tiny linear}}(t,x,\Omega)=\sum_{n=0}^{\infty}\sum_{l,k}\left\{A_{nlk}e^{-i\om_{nl} t}\,e_{nlk}(x,\Omega)+\bar A_{nlk}e^{i\om_{nl} t}\,\bar e_{nlk}(x,\Omega)\right\},
\eeq
where $A_{nlk}$ are arbitrary complex constants. 

It is well-known that attempting to correct the linearized solution (\ref{AdSlin}) perturbatively by building an expansion in powers of the overall amplitude in a way that satisfies (\ref{AdSwave}) fails to provide an adequate approximation. The reason lies in the emergence of {\it secular terms} in this expansion \cite{BR,CEV1,period} that grow in time and invalidate the expansion for initial data of amplitude $\eps$ on time scales of order $1/\eps^2$, precisely the scales that interest us here. The profusion of
secular terms is considerably enhanced by the presence of the resonances (\ref{resgen}). A variety of techniques exist to resum the secular terms \cite{FPU,CEV1,CEV2} and obtain an adequate approximation to the long-time weakly nonlinear dynamics of (\ref{AdSwave}). Rather than discussing here these resummation techniques, we shall phrase the derivation in the language of {\it time averaging} \cite{murdock}, which goes back to the classic works of Bogoliubov and Krylov, and will produce an effective resonant system completely equivalent to what one would have gotten from the secular term resummation as in \cite{FPU,CEV1,CEV2}. This derivation based on time-averaging \cite{CEV2} is more intuitive and compact.

Rather than naively expanding the solutions of (\ref{AdSwave}) in powers of the amplitude, we observe that, for small nonlinearities (small amplitude solutions), one may still treat the evolution as being of the form (\ref{AdSlin}) except that $A_{nkl}$ are no longer constant, but turn into slowly varying functions of time. In other words, small nonlinearities manifest themselves as slow modulations of the complex amplitudes of the linearized modes (which would be exactly constant by construction in a fully linearized theory). To express this idea mathematically, we perform a canonical transformation from $\phi$ and $\del\phi/\del t$ (which is related to the canonical momentum conjugate to $\phi$) to new complex dynamical variables $\al_{nkl}$ defined by
\begin{align}\label{AdSint}
&\phi(t,x,\Omega)=\eps \sum_{n=0}^{\infty}\sum_{l,k}\frac1{\sqrt{2\om_{nlk}}}\left\{\al_{nlk}(t)e^{-i\om_{nlk} t}\,e_{nlk}(x,\Omega)+\ab_{nlk}(t)e^{i\om_{nlk} t}\,\bar e_{nlk}(x,\Omega)\right\},\\
&\frac{\del\phi}{\del t}(t,x,\Omega)=-i\eps \sum_{n=0}^{\infty}\sum_{l,k}\sqrt{\frac{\om_{nlk}}2}\left\{\al_{nlk}(t)e^{-i\om_{nlk} t}\,e_{nlk}(x,\Omega)-\ab_{nlk}(t)e^{i\om_{nlk} t}\,\bar e_{nlk}(x,\Omega)\right\},\label{AdSint1}
\end{align}
where we have added an extra index $k$ to $\om_{nl}$ to keep track of which exact normal mode this frequency refers to, though the actual value of $\om$ given by (\ref{eq_eigenvalueAdS}) is of course $k$-independent.
Note that one should not think of the second line as a time derivative of the first. Rather, the two equations are independent and necessary to give a complete definition of the complex variable $\al_{nlk}$ in terms of real variables $\phi$ and $\del\phi/\del t$. One can think of this as a canonical transformation in the Hamiltonian formalism, and the new conjugate variables are $\al_{nlk}$ and $i\ab_{nlk}$.
Substituting these expressions in (\ref{AdSwave}) and projecting on $e_{nlk}$, one gets
\begin{align} \label{eq_preaveragingS3alpha}
&i\dsty\frac{d\al_{nlk}}{dt}=\eps^2\hspace{-1mm}\sum_{\begin{array}{l}\vspace{-7mm}\\\ssty n_1l_1k_1\vspace{-2mm}\\\ssty n_2l_2k_2\vspace{-2mm}\\\ssty n_3l_3k_3\\\vspace{-10mm}\end{array}}\int\limits_0^{\pi/2}dx\frac{\tan^{d-1}{x}}{\cos^2{x}} \int d\Omega_{d-1}\frac{\dsty e^{i\om_{nlk} t}\bar{e}_{nlk}}{\dsty 4\sqrt{\om_{nlk}\om_{n_1l_1k_1}\om_{n_2l_2k_2}\om_{n_3l_3k_3}}}\\
&\hspace{3.7cm}\times\left\{\al_{n_1l_1k_1}(t)e^{-i\om_{n_1l_1k_1} t}\,e_{n_1l_1k_1}(x,\Omega)+\ab_{n_1l_1k_1}(t)e^{i\om_{n_1l_1k_1} t}\,\bar e_{n_1l_1k_1}(x,\Omega)\right\}\nonumber\\
&\hspace{3.7cm}\times\left\{\al_{n_2l_2k_2}(t)e^{-i\om_{n_2l_2k_2} t}\,e_{n_2l_2k_2}(x,\Omega)+\ab_{n_2l_2k_2}(t)e^{i\om_{n_2l_2k_2} t}\,\bar e_{n_2l_2k_2}(x,\Omega)\right\}\rule{0mm}{5mm}\nonumber\\
&\hspace{3.7cm}\times\left\{\al_{n_3l_3k_3}(t)e^{-i\om_{n_3l_3k_3} t}\,e_{n_3l_3k_3}(x,\Omega)+\ab_{n_3l_3k_3}(t)e^{i\om_{n_3l_3k_3} t}\,\bar e_{n_3l_3k_3}(x,\Omega)\right\}\rule{0mm}{5mm}.\nonumber
\end{align}
At this point, the idea of time averaging comes in naturally. The time derivative of $\al_{nlk}$ is proportional to $\eps^2$ and hence small, so that $\al_{nlk}$ vary very slowly, on time scales $1/\eps^2$. At the same time, most terms on the right-hand side are oscillatory functions of $t$ with periods of order 1. It is natural to expect that the contributions of such oscillatory terms are negligible upon integration for sufficiently small $\eps$.
{\it Time averaging} or {\it resonant approximation} consists in precisely discarding all such oscillatory terms. Besides the general intuition that we have provided, rigorous proofs can be given that the resulting simplified dynamics approximates the original system arbitrarily well for sufficiently small $\eps$ on time scales $1/\eps^2$. For dynamical systems with a finite number of degrees of freedom, rigorous approximation results can be read off the textbook exposition \cite{murdock}. We are not aware of full-fledged proofs of this sort for nonlinear field equations in AdS, but they do exists in the mathematical literature \cite{GHT} for the closely related nonlinear Schr\"odinger equations in harmonic potentials, which arise as nonrelativistic limits of the AdS equations, see section~\ref{sec8}. Considerable numerical work with AdS systems further supports the validity of resonant approximation in this setting.

We must thus identify the terms on the right-hand side of (\ref{eq_preaveragingS3alpha}) that do not come with an explicit oscillatory time-dependent factor. The only way such terms may arise is through exact cancellations of the various oscillatory factors occurring in the products, which requires
\beq\label{ompm}
\om_{nlk}\pm\om_{n_1l_1k_1}\pm\om_{n_2l_2k_2}\pm\om_{n_3l_3k_3}=0,
\eeq
where the three plus-minus signs are independent. The highly resonant structure of the spectrum (\ref{eq_eigenvalueAdS}) leaves many options
for satisfying these relations.
One particular resonance that is present for any value of $\de$ is $\om_{nlk}+\om_{n_1l_1k_1}=\om_{n_2l_2k_2}+\om_{n_3l_3k_3}$ (and other similar relations obtained by permuting the subscripts 1, 2 and 3). By (\ref{eq_eigenvalueAdS}), this simply translates to $2n+l+2n_1+l_1=2n_2+l_2+2n_3+l_3$.
For special values of $\de$ (as in the case of massless fields), other resonances are theoretically possible, for example $\om_{nlk}=\om_{n_1l_1k_1}+\om_{n_2l_2k_2}+\om_{n_3l_3k_3}$. However, evaluation of the 
integrals over $x$ and $\Omega$ in (\ref{eq_preaveragingS3alpha}) leaves zero for these cases due to powerful {\it selection rules} to be discussed in more detail in section~\ref{sec4}. As a result,
only resonances of the form $\om_{nlk}+\om_{n_1l_1k_1}=\om_{n_2l_2k_2}+\om_{n_3l_3k_3}$ survive 
the time-averaging procedure, resulting in the following {\it resonant approximation} to (\ref{eq_preaveragingS3alpha}):
\beq
i\frac{d\al_{nlk}}{dt}=3\eps^2\hspace{-3mm}\sum_{\om+\om_1=\om_2+\om_3}\hspace{-3mm}C_{nlk,n_1l_1k_1,n_2l_2k_2,n_3l_3k_3}\,\ab_{n_1l_1k_1}\,\al_{n_2l_2k_2}\,\al_{n_3l_3k_3},
\label{reswavet}
\eeq
where $\om_{n_1l_1k_1}$ has been abbreviated as $\om_1$, and similarly for other modes. (The factor of 3 arises because, when expanding (\ref{eq_preaveragingS3alpha}), the complex conjugation could fall on either
$\al_{n_1l_1k_1}$, $\al_{n_2l_2k_2}$ or $\al_{n_2l_2k_2}$, and the three contributions equal each other due to the symmetries of $C$; this factor is, however, inconsequential and will be absorbed in a redefinition of time shortly.) The interaction coefficients $C$ capture the strength of nonlinear interactions of resonant mode quartets and are defined as
\beq
C_{nlk,n_1l_1k_1,n_2l_2k_2,n_3l_3k_3}=\int\limits_0^{\pi/2}dx\frac{\tan^{d-1}{x}}{\cos^2{x}} \int d\Omega_{d-1} \,\,\frac{\dsty \bar{e}_{nlk}\bar{e}_{n_1l_1k_1}e_{n_2l_2k_2}e_{n_3l_3k_3}}{\dsty 4\sqrt{\om_{nlk}\om_{n_1l_1k_1}\om_{n_2l_2k_2}\om_{n_3l_3k_3}}}.
\label{Cgen}
\eeq
By the nature of time-averaging, all explicit time dependencies have disappeared in (\ref{reswavet}) and
the evolution unfolds solely in terms of the {\it slow time} $\tau=3\eps^2 t$. One hence gets the final form of the resonant system for (\ref{AdSwave}):
\beq
i\,\frac{d\al_{nlk}}{d\tau}=\hspace{-3mm}\sum_{\om+\om_1=\om_2+\om_3}\hspace{-3mm}C_{nlk,n_1l_1k_1,n_2l_2k_2,n_3l_3k_3}\,\ab_{n_1l_1k_1}\,\al_{n_2l_2k_2}\,\al_{n_3l_3k_3},
\label{reswave}
\eeq
The small parameter $\eps$ no longer appears in the equation and has been absorbed in a redefinition of time. The resonant approximation thus captures in a universal form the weakly nonlinear dynamics with amplitudes $\eps$ on time-scales $1/\eps^2$, and a single solution of the resonant system corresponds to a family of approximate solutions of (\ref{AdSwave}) parametrized by $\eps$. Note that the derivation we have given
would remain essentially unchanged if one decided to replace $\phi^3$ on the right-hand side of (\ref{AdSwave})
by any other cubic nonlinearity, though some details (as the validity of selection rules for the resonant mode couplings) may have to be double-checked, and the expression for the interaction coefficients (\ref{Cgen}) will evidently change. In the language of the resonant system (\ref{reswave}), the form of nonlinearities is completely encoded in the expressions for the interaction coefficients $C$.

While it is not directly visible from the way everything has been derived here, each term on the right-hand side of (\ref{reswave}) corresponds to a particular secular term at order $\eps^3$ in the naive perturbative expansion of solutions to (\ref{AdSwave}) in powers of $\eps$. Derivations that make this relation explicit may be found in \cite{CEV1}. Thus, instead of saying that a particular interaction coefficient $C$ vanishes, one could equally well be talking about the absence of a particular secular term in the naive perturbative expansion. Such language has been used in parts of the literature on AdS perturbations, in particular, in relation to constructing time-periodic solutions.

The equation of motion (\ref{AdSwave}) is rotationally symmetric, and hence the interaction coefficients
(\ref{reswave}) must encode angular momentum conservation. In practice, it means that, if one introduces
the angular momentum projection on the polar axis $m(l,k)$ corresponding to the spherical harmonic labelled by $l$ and $k$, only terms on the right-hand side of (\ref{reswave}) satisfying $m(l,k)+m(l_1,k_1)=m(l_2,k_2)+m(l_3,k_3)$ will be actually present in the sum. This constraint is in addition
to the resonant constraint $\om+\om_1=\om_2+\om_3$ explicitly displayed in (\ref{AdSwave}).

As a matter of fact, to optimize the handling of such angular momentum conservation constraints
it would be logical to work in Hopf-like coordinates (see for instance \cite{hopf}) rather than in hyperspherical coordinates
(which is what we have done, and the standard formulation in the AdS literature). There are many potential
advantages in using Hopf-like coordinates for field equations in AdS \cite{J,P} as they by construction
introduce the maximal number of angular momentum components that can simultaneously have definite values, which is the number of Cartan generators of the rotation group, making the momentum conservation
as explicit as possible.

Apart from rotational symmetry, resonant system (\ref{reswave}) has a few further symmetries. First, all $\alpha$'s can be rotated by a common phase $e^{i\psi}$. Second, each $\alpha$ can be rotated with a frequency-dependent phase $\alpha_{nlk}\to \al_{nlk}e^{i \om_{nl}\psi}$. Third, all $\al$'s can be scaled by a common factor together with a rescaling of time
\beq
\al(\tau)\to \lambda\,\al(\lambda^2\tau).
\eeq
The first two symmetries generate conserved quantities by the Noether procedure, while the scaling symmetry does not (it respects the equations of motion, but not any variational principle from which the equations of motion descend).

The resonant dynamics of (\ref{reswave}) is defined on the full set of AdS normal modes depicted in fig.~\ref{figtower}, but it can be consistently truncated to many smaller and more manageable sets of modes. One such interesting truncation is to the maximally rotating modes (the modes of maximal angular momentum for a given frequency) treated in \cite{BEL}. Perhaps the most common truncation, however,
is imposing spherical symmetry on solutions of (\ref{AdSwave}), which is always possible on account of the rotational symmetry of this equation. In the language of normal modes, this simply amounts to discarding all modes with $l>0$. Since there is only one mode with $l=0$ for each $n$, one can simply re-label such modes as $\al_n$. Furthermore, the resonant condition $\om+\om_1=\om_2+\om_3$ becomes simply
$n+n_1=n_2+n_3$. As a result, one obtains a resonant system precisely of our prototypical form
(\ref{reseq}), with the interaction coefficients given specific values originating from (\ref{AdSwave}). One such system, corresponding to a conformally coupled scalar, will play a significant role further on in section~\ref{sec6}.

A technical comment is in order here. Two different conventions for the definition of $\al$ have been used in the literature. First, $\al$ can be introduced as in (\ref{AdSint}-\ref{AdSint1}) so that the left-hand side of (\ref{reswave}) is simply $id\al_{nlk}/d\tau$ and the interaction coefficients are as in (\ref{Cgen}).
Second, one may denote what we call $\al_{nlk}/\sqrt{2\om_{nlk}}$ as simply $\al_{nlk}$. In this case, the resonant equation becomes
\beq
i\,\om_{nlk}\,\frac{d\al_{nlk}}{d\tau}=\hspace{-3mm}\sum_{\om+\om_1=\om_2+\om_3}\hspace{-3mm}\tilde C_{nlk,n_1l_1k_1,n_2l_2k_2,n_3l_3k_3}\,\ab_{n_1l_1k_1}\,\al_{n_2l_2k_2}\,\al_{n_3l_3k_3},
\label{reswave2}
\eeq
while the interaction coefficients $\tilde C$ are given by the same formula as (\ref{Cgen})
but without the denominator $4\sqrt{\om_{nlk}\om_{n_1l_1k_1}\om_{n_2l_2k_2}\om_{n_3l_3k_3}}$.
If the evolution of the above equation is consistently truncated to a smaller set of modes labelled by a single integer, as described above, one obtains the following alternative form of (\ref{reseq}):
\beq
i\,\om_n\,\frac{d\al_n}{dt}=\sum_{n+m=k+l} \tilde C_{nmkl}\,\ab_m\al_k\al_l,
\label{reseq2}
\eeq
Both normalizations of $\al$, as in (\ref{reseq}) and (\ref{reseq2}), have found some uses and should be kept in mind. The $\tilde C$ form of the equations will be used, in particular, in the following exposition on gravitating perturbations.


\section{Gravitating perturbations in spherical symmetry}\label{sec3}

Having reviewed the nonlinear equations for non-gravitating fields in AdS background (\ref{AdSwave}) and effective analysis of their weakly nonlinear dynamics,
we can turn to the problem of gravitating perturbations in AdS. Here, due to the complexity of Einstein's equations, one must make choices
to keep the problem manageable.

The most straightforward approach is to impose spherical symmetry on the perturbations, so that the metric functions only depend on the radial variable and on time. This setup is what has underlied the analysis of \cite{BR} that has led historically to the formulation of the AdS instability conjecture, and has also been the basis for much of the subsequent work. We shall focus on this spherically symmetric setup here, and comment on the situation outside spherical symmetry, much less explored, in the next section.

While imposing spherical symmetry substantially simplifies Einstein's equations, it comes with a price. In spherical symmetry, gravity loses propagating
degrees of freedom, and the only vacuum solutions are static AdS-Schwarzschild black holes. To introduce nontrivial dynamics to the problem, one must add matter to the system, which we shall choose to be a massless scalar field following \cite{BR}. In this formulation, due to spherical symmetry,
the metric components are completely expressed by the constraint equations through the configuration of the scalar field on each fixed time slice, and
the dynamics can be recast as an effective wave equation for the scalar field, making the problem very similar to the non-gravitating case of the previous section.

The basic setup is the following system of coupled Einstein's equations and wave equation for a real scalar field $\phi$:
\begin{equation}
G_{\mu\nu}+\Lambda g_{\mu\nu}=8\pi G\left(\partial_{\mu}\phi\partial_{\nu}\phi-\frac{1}{2}g_{\mu\nu}(\partial\phi)^{2}\right)
\qquad\text{and}\qquad
\Box\phi=0.
\end{equation}
(It is convenient to set $\Lambda=-{d(d-1)}/{2}$ and $8\pi G=d-1$ by a choice of units, which we shall assume from now on.)
A general spherically symmetric geometry can be parametrized by two functions $A(x,t)$ and $\delta(x,t)$ as
\begin{equation}\label{AdSanstz}
ds^{2}=\frac{1}{\cos^{2}x}\left(\frac{dx^{2}}{A}-Ae^{-2\delta}dt^{2}+\sin^{2}x\,d\Omega_{d-1}^{2}\right).
\end{equation}
This ansatz is adapted to asymptotically AdS spacetimes, $t$ runs over the real axis and $x$ runs from 0 to $\pi/2$. 
The AdS metric (\ref{adsmetric}) is recovered by setting $A=1$ and $\de=0$.
We assume that scalar field is spherically symmetric $\phi=\phi(x,t)$ and introduce, following \cite{BR}, $\Phi\equiv\phi'$ and $\Pi\equiv A^{-1}e^{\delta}\dot{\phi}$ (dots and primes stand for the derivatives with respect to $t$ and $x$). We also introduce two specific trigonometric functions that will occur often in the formulas below:
\begin{equation}
\mu(x)\equiv(\tan x)^{d-1}
\qquad\text{and}\qquad
\nu(x)\equiv\frac{\sin x\cos x}{(\tan x)^{d-1}}.
\label{munu}
\end{equation}
With these preliminaries, the equations of motion become \cite{BR}
\begin{align}
&\hspace{2cm}\dot{\Phi}=\left(Ae^{-\delta}\Pi\right)',\qquad
\dot{\Pi}=\frac{1}{\mu}\left(\mu Ae^{-\delta}\Phi\right)', \label{EOMphi} \\
&A'=\frac{\nu'}{\nu}\left(A-1\right)-\mu\nu\left(\Phi^{2}+\Pi^{2}\right)A,\qquad
\delta'=-\mu\nu\left(\Phi^{2}+\Pi^{2}\right),\label{eqn:EOMConstraint}
\end{align}
If $\phi=0$, and hence $\Phi=\Pi=0$, the only solution, recovered through a simple integration $\int dA/(A-1)=\int d\nu/\nu$, is
an AdS-Schwarzschild black hole $A(x,t)=1-M\nu(x)$, $\delta(x,t)=0$, as mentioned above.

The ansatz (\ref{AdSanstz}) still has some gauge redundancy in the form of arbitrary time 
reparametri\-zation. This lets one set $\de$ to any chosen function of time at one prescribed value of $x$.
There are two common choices that have appeared in the literature: the `interior gauge' $\de(x=0)=0$, corresponding to measuring the time at the center of spherical symmetry, and the `boundary gauge'
$\de(x=\pi/2)=0$ corresponding to measuring the time at the AdS boundary. In the interior gauge,
the collapse is seen more explicitly, and it is a natural choice for numerical simulations as in \cite{BR}.
The boundary gauge has the advantage that the resonant system we are about to construct is explicitly Hamiltonian. In any case, considerations in the two gauges are straightforward to relate, though the explicit transformations involve the entire history of the system, see \cite{CEV2}.

Dynamics of the system (\ref{EOMphi}-\ref{eqn:EOMConstraint}) has been extensively considered and
played a crucial role in the formulation of the AdS instability conjecture in \cite{BR}.
The principal observation about its dynamics is that, for some initial perturbation profiles in terms of the initial conditions for $\phi$, the system ends up forming a black hole at the origin after multiple bounces of the wave front between the AdS interior and boundary, no matter how much one scales down the amplitude
of the initial perturbation. For initial perturbations of amplitude $\eps$, the time necessary for collapse (and the number of wave front bounces) scales as $1/\eps^2$. Formation of a small black hole at the end of this process (given the small amount of available total energy) requires transfer of energy towards short-wavelength modes, in other words, a turbulent cascade. Later on, initial perturbation profiles were discovered that do not apparently lead to collapse \cite{BLL}. One may think of `islands of stability' \cite{period} in an ocean of instability. 
The conjecture still remains that an open set of initial profiles exists for which collapse happens no matter how much the amplitude is scaled down.

The relevant dynamics underlying the AdS instability happens on time scales $1/\eps^2$ for perturbations
of amplitude $\eps$, which is precisely the domain of validity of the resonant approximation to (\ref{EOMphi}-\ref{eqn:EOMConstraint}), making it very natural to appeal to the resonant approximation 
in this context. The idea to pursue this approach was briefly flashed already in \cite{BR}, but it took a few
years before it got implemented in practice. The resonant approximation (under the name of the `two-time framework') was introduced to studies of (\ref{EOMphi}-\ref{eqn:EOMConstraint}) in \cite{FPU}, mostly
as an alternative approach to numerical simulations. It was then given a fully analytic formulation in \cite{CEV1,CEV2}. This resonant system has been further numerically analyzed in \cite{BMR}, displaying a turbulent cascade suggestive of black hole formation, and lending further credence to the AdS instability conjecture. Below, we will briefly review the resulting resonant system, giving a heuristic sketch of how the relevant structures emerge (different from the original derivations of \cite{CEV1}) and referring the reader to \cite{CEV1,CEV2} for further details.

As has been already remarked, gravity has no propagating degrees of freedom in spherical symmetry,
which in practice means that $A$ and $\de$ can be `integrated out' from (\ref{EOMphi}-\ref{eqn:EOMConstraint}), that is, expressed through the configuration of matter on each $t=\cnst$ slice. 
This procedure leaves a complicated effective nonlinear wave equation for $\phi$ that can be treated with the methods of the previous section. To implement this procedure, one rewrites (\ref{eqn:EOMConstraint}) as
\begin{equation}\label{eqA}
1-A=e^{\delta}\nu\int_{0}^{x}\text{d}y\left(\Phi^{2}+\Pi^{2}\right)e^{-\delta}\mu
\end{equation}
and
\begin{equation}\label{eqde}
\delta=\int_{x}^{\pi/2}\text{d}y\left(\Phi^{2}+\Pi^{2}\right)\mu\nu,
\end{equation}
where the boundary gauge is chosen for $\de$, though this can be easily changed.
A solution to these equations can be written as an expansion in powers of $\phi$, with the first nontrivial term being of order $\phi^2$. This first nontrivial correction $A=1+O(\phi^2)$, $\de=O(\phi^2)$ can then be substituted into the scalar field equation of motion $\Box\phi=0$. Since $A=1,\de=0$ corresponds to pure AdS, the result can be rewritten as
\beq
\Box_{AdS}\phi=Q[\phi],
\label{effwave}
\eeq
where $\Box_{AdS}$ is the d'Alambertian of the metric (\ref{adsmetric}), and $Q$ is cubic in $\phi$ and expressed through $\phi'$ and $\dot\phi$ (it is local in time but involves integrals over $x$). We quote an explicit effective action due to \cite{CEV2} from which this effective wave equation can be derived:
\begin{equation}
S=\int\text{d}t\int\text{d}x\left\{(\phi'^{2}-\dot{\phi}^{2})\mu-(\phi'^{2}+\dot{\phi}^{2})\mu\nu\int_{0}^{x}\text{d}y\,(\phi'^{2}+\dot{\phi}^{2})\mu\right\}.
\end{equation}

Once the problem has been recast as (\ref{effwave}), the analysis of the previous section applies essentially verbatim, though the expressions are considerably more bulky. Remembering that we have imposed spherical symmetry, (\ref{AdSint}) is replaced by a simplified expansion, only involving the (real-valued)
zero angular momentum mode functions $e_n(x)\equiv e_{n00}(x,\Omega)$, which can be read off (\ref{eq_modefunctionAdS}) remembering that $Y_{00}=1$. Explicitly, this expansion is
\beq
\phi(x,t)=\sum_{n=0}^\infty \left(\al_n(t)e^{-i\om_n t}+\ab_n(t)e^{i\om_n t}\right)\,e_n(x),
\eeq
where $\om_n=d+2n$ and we have removed the frequency-dependent denominators from (\ref{AdSint}) to match the conventions of \cite{CEV1,CEV2}. Re-expressing the dynamics in terms of $\al_n$ and applying time-averaging, after considerable algebraic work, one obtains the corresponding resonant system fitting the general pattern of (\ref{reseq2}), which can be expressed, as in \cite{CEV2}, in the following form:
\begin{equation}
\label{eqn:RG2}
i \omega_{l}\frac{d\alpha_{l}}{d\tau}=T_{l}|\alpha_{l}|^{2}\alpha_{l}+\sum_{i}^{i\,\neq\,l}R_{il}|\alpha_{i}|^{2}\alpha_{l}+\underbrace{\sum_{i}^{\{i,j\}}\sum_{j}^{\neq}\sum_{k}^{\{k,l\}}}_{\omega_{i}+\omega_{j}=\omega_{k}+\omega_{l}}S_{ijkl}\alpha_{i}\alpha_{j}\bar{\alpha}_{k}.
\end{equation}
Here, we have separated the interaction coefficients $C$ into three different objects $T$, $R$ and $S$, according to the number of coincident indices, since the algebraic expressions for these three objects are different. The explicit form of the interaction coefficients expressed through the mode functions is \cite{CEV2}
\begin{align}
T_{l}=&\frac{1}{2}\omega_{l}^{2}X_{llll}+\frac{3}{2}Y_{llll}+2\omega_{l}^{4}W_{llll}+2\omega_{l}^{2}\tilde W_{llll}-\omega_{l}^{2}(A_{ll}+\omega_{l}^{2}V_{ll}), \label{Tdef}\\
R_{il}=&
\frac{1}{2}\left(\frac{\omega_{i}^{2}+\omega_{l}^{2}}{\omega_{l}^{2}-\omega_{i}^{2}}\right)\left(\omega_{l}^{2}X_{illi}-\omega_{i}^{2}X_{liil}\right)+2\left(\frac{\omega_{l}^{2}Y_{ilil}-\omega_{i}^{2}Y_{lili}}{\omega_{l}^{2}-\omega_{i}^{2}}\right)+\left(\frac{\omega_{i}^{2}\omega_{l}^{2}}{\omega_{l}^{2}-\omega_{i}^{2}}\right)\left(X_{illi}-X_{lili}\right) \nonumber \\
&+\frac{1}{2}(Y_{iill}+Y_{llii})+\omega_{i}^{2}\omega_{l}^{2}(W_{llii}+W_{iill})+\omega_{i}^{2}\tilde W_{llii}+\omega_{l}^{2}\tilde W_{iill}-\omega_{l}^{2}(A_{ii}+\omega_{i}^{2}V_{ii}),\\
S_{ijkl}=&
-\frac{1}{4}\left(\frac{1}{\omega_{i}+\omega_{j}}+\frac{1}{\omega_{i}-\omega_{k}}+\frac{1}{\omega_{j}-\omega_{k}}\right)(\omega_{i}\omega_{j}\omega_{k}X_{lijk}-\omega_{l}Y_{iljk}) \nonumber \\
&-\frac{1}{4}\left(\frac{1}{\omega_{i}+\omega_{j}}+\frac{1}{\omega_{i}-\omega_{k}}-\frac{1}{\omega_{j}-\omega_{k}}\right)(\omega_{j}\omega_{k}\omega_{l}X_{ijkl}-\omega_{i}Y_{jikl}) \nonumber \\
&-\frac{1}{4}\left(\frac{1}{\omega_{i}+\omega_{j}}-\frac{1}{\omega_{i}-\omega_{k}}+\frac{1}{\omega_{j}-\omega_{k}}\right)(\omega_{i}\omega_{k}\omega_{l}X_{jikl}-\omega_{j}Y_{ijkl}) \nonumber \\
&-\frac{1}{4}\left(\frac{1}{\omega_{i}+\omega_{j}}-\frac{1}{\omega_{i}-\omega_{k}}-\frac{1}{\omega_{j}-\omega_{k}}\right)(\omega_{i}\omega_{j}\omega_{l}X_{kijl}-\omega_{k}Y_{ikjl}),
\label{Sdef}
\end{align}
where $X$, $Y$, $W$, $\tilde W$, $V$ and $A$ are the following integrals of quartic combinations of the mode functions and their derivatives:
\begin{align*}
&X_{ijkl}=\int_{0}^{\pi/2}\text{d}x\,e'_{i}(x)e_{j}(x)e_{k}(x)e_{l}(x)(\mu(x))^{2}\nu(x), \\
&Y_{ijkl}=\int_{0}^{\pi/2}\text{d}x\,e'_{i}(x)e_{j}(x)e'_{k}(x)e'_{l}(x)(\mu(x))^{2}\nu(x), \\
&W_{ijkl}=\int_{0}^{\pi/2}\text{d}x\,e_{i}(x)e_{j}(x)\mu(x)\nu(x)\int_{0}^{x}\text{d}y\,e_{k}(y)e_{l}(y)\mu(y), \\
&\tilde W_{ijkl}=\int_{0}^{\pi/2}\text{d}x\,e'_{i}(x)e'_{j}(x)\mu(x)\nu(x)\int_{0}^{x}\text{d}y\,e_{k}(y)e_{l}(y)\mu(y), \\
&V_{ij}=\int_{0}^{\pi/2}\text{d}x\,e_{i}(x)e_{j}(x)\mu(x)\nu(x), \qquad
A_{ij}=\int_{0}^{\pi/2}\text{d}x\,e'_{i}(x)e'_{j}(x)\mu(x)\nu(x).
\end{align*}
These expressions are for the interior gauge $\de(x=0)=0$. They can be straightforwardly converted to the corresponding resonant system in the boundary gauge, as explained in \cite{CEV2}. Note that these complicated expressions replace the simple form of the interaction coefficients for the non-gravitating scalar field (\ref{Cgen}), while being only valid in spherical symmetry.

We note, importantly, that the right-hand side of (\ref{eqn:RG2}) does not contain any terms of the form $\alpha_k\alpha_m\alpha_n$ or $\ab_k\ab_m\al_n$, which could have been present due to the resonances $\om_l=\om_k+\om_m+\om_n$ and $\om_l+\om_k+\om_m=\om_n$.
The point, just like in the previous section, is that the corresponding interaction coefficients vanish, as proved in \cite{CEV1}. These AdS selection rules have immediate dynamical consequences. First of all, single-mode solutions are stationary (if only one mode is nonzero in the initial state, no other modes will ever get excited). These modes have been seen as the seeds of islands of stability \cite{period}, see also \cite{BLL,FPU,CEV1,CEV2,dual,GMLL}. The selection rules furthermore enhance \cite{CEV2} the set of conserved quantities of (\ref{eqn:RG2}), ensuring that
\beq\label{sphercons}
N=\sum_{n=0}^\infty \om_n|\al_n|^2\quad\mbox{and}\quad E=\sum_{n=1}^\infty n\om_n|\al_n|^2
\eeq
are both conserved (only the second quantity would have been conserved if the selection rules were not in operation). This doublet of conservation laws
guarantees that turbulent transfer must take place in the form of dual cascades \cite{dual}. We shall return to the topic of selection rules in section~\ref{sec4}.


\section{Gravitational perturbations outside spherical symmetry}\label{sec3bis}

There is a number of reasons one might want to extend the considerations of the previous section
outside spherical symmetry, which is in full generality a rather formidable task, far from being accomplished. First, the question of the fate of general (rather than spherically symmetric)
perturbations is of interest in its own right. Second, understanding such general perturbation will provide
a better and more detailed contact with the AdS/CFT considerations (where imposing spherical symmetry is not very natural). Third, dynamics is impossible in spherical symmetry without matter added to the theory, and that may make one wonder whether the instability is driven by this matter content rather than being of gravitational nature.

The last concern can in fact be addressed without going to the full non-spherically-symmetric dynamics,
in a way that shows explicitly the gravitational nature of the instability. The idea is to specialize to AdS$_5$ and apply squashing \cite{BR2,squashed} to the 3-sphere appearing in the ansatz (\ref{AdSanstz}), which yields a Bianchi IX space instead of the 3-sphere. This breaks spherical symmetry, hence making purely gravitational vacuum dynamics possible, but still ensures that all the unknown metric functions depend only on time and the radial coordinate, keeping the problem manageable. The consistency of the ansatz is guaranteed because it can be enforced by imposing the symmetries of the Bianchi IX space on the corresponding spatial directions. The idea goes back to a similar treatment of asymptotically Minkowski solutions in \cite{BChSch}. 

The metric ansatz is then \cite{BR2,squashed}
\begin{equation}\label{AdSanstzsquash}
ds^{2}=\frac{1}{\cos^{2}x}\left(\frac{dx^{2}}{A}-Ae^{-2\delta}dt^{2}+\sin^{2}x\left[e^{2B}(\s_1^2+\s_2^2)+e^{-4B}\s_3^2\right]\right),
\end{equation}
where $\s_1+i\s_2=e^{\i\psi}(\cos\te d\varphi+ id\te)$ and $\s_3=d\psi-\sin\te d\varphi$. There are no extra matter fields and one is solving the vacuum Einstein equations with a negative cosmological constant for metrics of the form (\ref{AdSanstzsquash}), with $B$, $A$, $\de$ being functions of $t$ and $x$. In the resulting dynamics, $B$ takes on a role very similar to the role of the scalar field $\phi$ in the previous section, despite its purely gravitational origin. One observes the same range of phenomena, namely, black hole formation in numerical simulations starting from small initial data \cite{BR2} in the spirit of the AdS instability conjecture. Furthermore, one can analyze the dynamics in terms of the resonant approximation along the lines of \cite{CEV1,CEV2} obtaining a resonant system of a very similar structure \cite{squashed}. This makes the point that AdS instability in no way relies on the presence of the matter fields and can be realized with purely gravitational degrees of freedom.

As one moves away from the ansatz (\ref{AdSanstzsquash}), the situation becomes forbiddingly complicated. A number of partial treatments exist
in the literature \cite{nonspher1a,nonspher1b,nonspher2,nonspher3,nonspher4,nonspher5,nonspher6,nonspher7,highpert}, but they mostly focus on evaluating 
the interaction coefficients involving specific mode quartets and analyzing their properties. These treatments typically rely on the Regge-Wheeler formalism \cite{RW,Nollert} and the Ishibashi-Kodama construction \cite{IK}. Guidelines for a systematic analysis of higher-order perturbation theory outside spherical symmetry have been spelled out in \cite{highpert}. However, to this day, construction of a full resonant system for the gravitational sector
analogous to the resonant system (\ref{reswave}) for a probe scalar field has remained elusive.
(Recent work on numerical simulations of Einstein's equations in asymptotically AdS spacetimes outside spherical symmetry can be found in \cite{BFKR,BFR}.)

One particular subtlety that must be mentioned is that Einstein's equations involve quadratic nonlinearities, and yet the structure of the resonant approximation
outside spherical symmetry is expected to be characteristic of cubic nonlinearities, as in (\ref{reswave}). The reason is that quadratic nonlinearities in Einstein's equations do not generate secular terms, and hence their effect remains uniformly bounded on long time-scales. Secular terms only appear at the second nontrivial order of the perturbation theory, and the corresponding resonant approximation will involve a cubic resonant system similar to (\ref{reswave}). This vanishing of quadratic secular term, which is again a form of AdS selection rules, has been proved for the squashed sphere ansatz (\ref{AdSanstzsquash}) in \cite{squashed} and observed in practice in the numerical work for general perturbations outside spherical symmetry \cite{nonspher2,nonspher4,nonspher7}. Similarly, there are indications that selection rules for cubic terms in the resonant approximation are in place, similar to the selection rules described for probe scalars under (\ref{ompm}). They have been proved in \cite{squashed} within the squashed ansatz, and remain an empirical observation for the general case.

The bulk of the work outside spherical symmetry has focused so far not on constructing the resonant system to study the weakly nonlinear energy transfer, which is too complicated, but rather on analyzing naive perturbative expansions in powers of the perturbation amplitudes and studying secular terms arising in these expansions. In such treatments, an essential first step is to show that irremovable secular terms exist in the expansion. The notion of irremovable secular terms comes from the Poincar\'e-Lindstedt method of absorbing secular terms into small shifts of frequencies of the normal modes \cite{murdock}. This treatment is possible for any secular terms coming from the trivial four-mode resonances that correspond to the tautological resonant relations $\om_1+\om_2=\om_1+\om_2$; note that such four-mode resonances are present even for systems that do not possess any genuine resonant relations of the form (\ref{resgen}). These secular terms can always be absorbed into frequency shifts and are therefore called removable, while all other secular terms are irremovable. Irremovable secular terms cannot be absorbed into frequency shifts and indicate that significant energy transfer will take place between the normal modes on long time-scales. The presence of irremovable secular terms for gravitational AdS perturbations outside spherical symmetry was established in \cite{nonspher1a}. This guarantees that significant energy transfer takes place, though it does not guarantee that this transfer will be turbulent. The latter question requires actually controlling the form of the energy transfer by (resumming the secular terms and) working with the resonant approximation. This remains a topic for future research.

Another interesting question that can be addressed within naive perturbation theory is the existence
of time-periodic solutions of Einstein's equations in asymptotically AdS spacetimes (a prototype treatment of this problem in spherical symmetry is \cite{period}). The point is that while, in general, irremovable secular terms are present in the naive perturbative expansions for AdS perturbations, it is possible to fine-tune the  initial conditions in such a way that these secular terms disappear. Then, no significant energy transfer between the normal modes occurs, and one can resum the remaining removable secular terms into
Poincar\'e-Lindstedt frequency shifts. This construction results in special time-periodic solutions of the original
gravitational equations of motion, and it has been developed in  \cite{nonspher1a,nonspher1b,nonspher2,nonspher4}. 
At lowest order, this procedure can be recast in the language of the resonant system that suggestively
brings in the structure of a tensor eigenvalue problem \cite{tensor1, tensor2}. Since the resonant system
for gravitational perturbations in AdS is not available yet, though it is expected to take a form similar to (\ref{reswave}), we first phrase our discussion in terms of constructing time-periodic solutions of the cubic Klein-Gordon equation (\ref{AdSwave}), which translate into stationary solutions of the corresponding resonant system (\ref{reswave}) of the form
\beq\label{alstat}
\al_{nlk}(\tau)=e^{-i\lambda\tau}A_{nlk}, 
\eeq
where $A_{nlk}$ are time-independent.
We furthermore focus, as in \cite{nonspher1a,nonspher1b,nonspher2,nonspher4}, on solutions of this form for which all nonzero $\al_{nlk}(0)$ belong to the same frequency level $\om_{nl}=\om$. In this case, the structure of the resonant system (\ref{reswave}) guarantees that no $\al_{nlk}$ outside this frequency level ever get excited, and hence the dynamics can be consistently truncated to this frequency levels. The resonant condition degenerates to $\om+\om=\om+\om$ and is hence automatically satisfied and can be discarded. Finally, relabelling the different modes within $\al_{nlk}$ within the frequency level $\om_{nl}=\om$ as $\al_J$ (this index $J$ takes a finite number of distinct values for any given frequency level), the resonant system (\ref{reswave}) for stationary solutions of the form (\ref{alstat}) can be written as
\beq\label{hermtens}
\lambda A_J =\sum_{KLM} C_{JKLM}\bar A_K A_L A_M.
\eeq
Such equations are known as tensor eigenvalue problems \cite{tensor1,tensor2}. The presence of complex conjugation in $\bar A_K$ makes the equation slightly different, in fact, from the common definition of tensor eigenvalue problems discussed in \cite{tensor1,tensor2}. It is natural to call (\ref{hermtens}) a Hermitian tensor eigenvalue problem, since it naturally generalizes eigenvalue problems for Hermitian matrices.
It is easy to prove that $\lambda$ is always real: one should multiply both sides by $\bar A_J$ and sum over $J$ remembering the symmetries of $C$:
\beq
C_{n_1l_1k_1n_2l_2k_2n_3l_3k_3n_4l_4k_4}=C_{n_2l_2k_2n_1l_1k_1n_3l_3k_3n_4l_4k_4}=\bar C_{n_3l_3k_3n_4l_4k_4n_1l_1k_1n_2l_2k_2}.
\eeq 

Hermitian tensor eigenvalue problems naturally arise in the analysis of stationary solutions to resonant systems,
and they have been discussed informally in this context on a number of occasions following the publication of \cite{nonspher2}. More recently, these problems have been formally defined in mathematical works \cite{htensor1,htensor2} driven by rather different motivations. In contrast to the more conventional tensor eigenvalue problems \cite{tensor1,tensor2} that lack complex conjugation operations in the defining equation, very little is known about the Hermitian tensor eigenvalue and eigenvector properties. (The number of conventional tensor eigenvalues is known to be exponentially large in the number of dimensions.)

The equations defining the lowest nontrivial order time-periodic solutions for gravitational AdS perturbations
outside spherical symmetry, as those in \cite{nonspher2}, can be naturally recast\footnote{I thank Andrzej Rostworowski, Athanasios Chatzikaleas and Zbigniew B\l{}ocki for discussions on this form of the equations, and on counting the number of solutions in low-dimensional cases.} in the form of the tensor
eigenvalue problem (\ref{hermtens}). It has been conjectured in \cite{nonspher2}, based on the analysis of a few low-dimensional cases, that the number of solutions for gravitational perturbations is always the same as the number of dimensions of the corresponding single frequency subspace. This number is expected to be low compared to generic numbers of Hermitian tensor eigenvalues (though no systematic counting for such eigenvalues is available at the moment). The corresponding problem for time-periodic solutions of the cubic Klein-Gordon equation in AdS has also been considered \cite{M}, but it does not display a similar simple relation between the number of time-periodic solutions and the number of modes of a given frequency. One may hope to shed further light on the problem by recasting it in Hopf-like coordinates, already mentioned at the end of section~\ref{sec2}, but only initial efforts have been made in that direction \cite{S}.


\section{Selection rules}\label{sec4}

As has been highlighted in the previous sections, the AdS selection rules, namely the vanishing of some mode coupling coefficients, play a significant role in shaping the corresponding resonant dynamics. Were these mode couplings nonzero, they would have contributed extra terms on the right-hand side of the resonant system, upsetting a number of patterns in its dynamics. Some of the consequences \cite{CEV1,CEV2,dual} of the selection rules are the presence of the first of the two conservation laws (\ref{sphercons}), the fact that initially exciting only one frequency level does not lead to energy transfer outside this frequency level within the resonant approximation, as well as implications of the latter fact for islands of stability \cite{period} centered on single-mode solutions.

There are various levels of sophistication for the treatment of selection rules in different cases. For gravitational perturbations outside spherical symmetry,
they are only known from numerical evaluation of specific mode couplings, typically in a form where these mode couplings manifest themselves as coefficients of specific secular terms \cite{nonspher1a,nonspher1b,nonspher2,nonspher4,nonspher7}. These secular terms could be nonzero on the basis of resonant relations between the mode frequencies, but are in fact zero. For probe scalar fields treated in section~\ref{sec2}, the selection rules can be proved completely explicitly based on the properties of Jacobi polynomials, which are defined as a system of orthogonal polynomials and hence an integral of such a polynomial multiplied by any polynomial of a lower degree is zero by orthogonality. This orthogonality structure has been used to prove the selection rules for the cubic Klein-Gordon equation in AdS in \cite{Yang}. Similar (though more involved) proofs exist for gravitating scalars in spherical symmetry in \cite{CEV1} and for gravitational dynamics within the squashed sphere ansatz in \cite{squashed}.

While the proofs based on orthogonality of Jacobi polynomials are efficient and `hands-on,' it is tempting, in view of the importance of the selection rules,
to develop a deeper understanding of their origin based on some symmetry principles. While no such picture has been given for the selection rules
in the presence of gravitational interactions, the analysis of \cite{EN} provides this symmetry-based picture for the selection rules of the cubic Klein-Gordon equation in AdS, expressing them as a consequence of AdS isometries. We shall briefly review this symmetry-based derivation, referring the readers to \cite{EN} for technical details.

For concreteness, we specialize to (\ref{AdSwave}) for a massless scalar field whose normal mode frequencies
\beq
\om_{nlk}=d+2n+l
\eeq
are all integers. In this situation, when applying the resonant approximation to (\ref{eq_preaveragingS3alpha}), there are in principle resonances of the form $\om_{nlk}=\om_{n_1l_1k_1}+\om_{n_2l_2k_2}+\om_{n_3l_3k_3}$, which should contribute terms of proportional to $\al_{n_1l_1k_1}\al_{n_2l_2k_2}\al_{n_3l_3k_3}$ to the right-hand side of (\ref{reswavet}). The coefficients of such terms involve
\beq
Q_{nlk,n_1l_1k_1,n_2l_2k_2,n_3l_3k_3}=\int\limits_0^{\pi/2}dx\frac{\tan^{d-1}{x}}{\cos^2{x}} \int d\Omega_{d-1} \,\,\frac{\dsty \bar{e}_{nlk}{e}_{n_1l_1k_1}e_{n_2l_2k_2}e_{n_3l_3k_3}}{\dsty 4\sqrt{\om_{nlk}\om_{n_1l_1k_1}\om_{n_2l_2k_2}\om_{n_3l_3k_3}}}.
\label{Cvanish}
\eeq
Our purpose is to prove that these coefficients vanish. We shall replace the subscripts $nlk$, $n_1l_1k_1$,
$n_2l_2k_2$ and $n_3l_3k_3$ by 1, 2, 3 and 4 in the following discussion to keep the notation compact (since the entire discussion can be phrased for the four specific chosen mode functions, without referring to other mode functions, we may call them by any names we choose).

To make the operation of the AdS isometries more manifest, it is natural to extend the integrals over a single spatial slice in (\ref{Cvanish}) to the entire AdS hyperboloid (\ref{embedding}). To this end we simply define
\beq
\psi_J(t,x,\Omega)= e^{i\om_J t} e_J(x,\Omega).
\eeq
Since the frequencies $\om_J$ are integers, these functions are single-valued and continuous on the hyperboloid (\ref{embedding}). If one expresses $e_J$ through $\psi_J$ in (\ref{Cvanish}), all the oscillatory factors simply cancel out on account of the resonance condition $\om_1=\om_2+\om_3+\om_4$. Hence we can simply append an integral over $t$ from 0 to $2\pi$ and divide by $2\pi$ without changing the expression. 
Furthermore, $dt\,dx\,d\Omega\,\tan^{d-1}x/\cos^2x$ is the invariant integration measure of the metric (\ref{adsmetric}), and hence
\beq\label{C1234}
Q_{1234}\sim\int_{AdS} \bar\psi_1 \psi_2\psi_3\psi_4,
\eeq
where $\int_{AdS}$ simply means integrating over the entire hyperboloid (\ref{embedding}) with the invariant measure.

A crucial ingredient of the selection rule proof is that the entire set of $\psi$-functions corresponding to the AdS mode functions (\ref{eq_modefunctionAdS}) can be constructed by applying certain AdS isometry generators to the ground state $\psi$-function, which is $\psi_0\equiv e^{idt}e_{000}(x,\Omega)$. These generators, which we call $L_{i+}$ with $i$ running from $1$ to $d$ are analogous to Lorentz boosts in Minkowski space, and come together with their Hermitian conjugates $L_{i-}$ that annihilate $\psi_0$:
\beq
L_{i-}\psi_0=0.
\eeq
Any mode function can be obtained by repeatedly acting on $\psi_0$ with $L_{i+}$, which yields
\beq\label{raising}
\psi=L_{i_1+}\cdots L_{i_N+} \psi_0.
\eeq
The time dependence of this function is a simple oscillation $e^{i\om t}$ with frequency $\om=d+N$.
The set of functions obtained in this manner is complete. Indeed, as different $L_{i+}$ commute with each other, the above construct is a fully symmetric rank $N$ tensor in $d$ dimensions, which is precisely the total number of components of all the multiplets at level $N$ above the ground state in fig.~\ref{figtower}. Thus, in particular, mode functions $e_{nlk}$ of (\ref{eq_modefunctionAdS}) can be expressed through these functions by a linear change of basis. It thus suffices to prove the selection rules in the basis constructed as in (\ref{raising}).

The isometry generators $L_{i+}$ possess a few essential properties. First, as all isometries, they are first order differential operators. Second, their Hermitian conjugates (with respect to the inner product defined by invariant AdS integrals) are isometry generators $L_{i-}$ that act as lowering operators on (\ref{raising}). This can be seen from the commutation relations of the $SO(d,2)$ isometry algebra \cite{EN}. In particular, acting with more than $N$ such operators on (\ref{raising}) always annihilates it. With all of the above, we equivalently rewrite (\ref{C1234}) as
\beq\label{CL1234}
Q_{1234}\sim\int_{AdS} \overline{L_{i_1+}\cdots L_{i_{N}+}\psi_0} \left\{(L_{j_1+}\cdots L_{j_{P}+}\psi_0)(L_{k_1+}\cdots L_{k_{R}+}\psi_0)(L_{l_1+}\cdots L_{l_{S}+}\psi_0)\right\},
\eeq
We can use Hermitian conjugations to take the operators $L_{i+}$ off the first factor, turn them into $L_{i-}$ and act with them on the content of the curly brackets. Being first order differential operators, they will simply distribute in different ways among the three factors inside the curly brackets. But the resonance condition $\om_1=\om_2+\om_3+\om_4$ implies that $N=d+P+R+S$, hence $N$ is too large and there are two many $L_{i-}$ operators generated in this fashion. No matter how they are distributed among the three factors inside the curly brackets of (\ref{CL1234}), at least one of the factors will be annihilated. Hence, $Q$ vanishes and the selection rule has been proved.

It is wise to highlight a particular attractive aspect of the representation (\ref{C1234}). We started with the finite-time evolution of the dynamical equation (\ref{AdSwave}) written in a particular coordinatization of the AdS space, which implies a time-slicing prescription. The general covariance is far from manifest in this formulation. Nonetheless, after converting the original equations to the resonant approximation and extracting the interaction coefficients, we were able to rewrite them in a manifestly covariant form, completely independent of the AdS coordinatization. Similar rewriting is easily accomplished for the nonvanishing interaction coefficients (\ref{Cgen}). It is natural to expect that similar re-structuring of the interaction coefficients of the more complicated gravitational problem could be of great use, but it has not been devised as yet.

Expressions of the form (\ref{C1234}) display the same structure as what is known as the
`tree level amplitudes' in quantum field theory, namely, the nonlinear part of the action with
the mode functions substituted in place of the fields (as far as the simple case of a scalar with quartic contact interactions is concerned). Indeed, if we replace AdS with Minkowski space in (\ref{C1234})
and choose $\psi_I$ as the corresponding plane wave mode functions $\psi_I=e^{ik_Ix}$, (\ref{C1234}) turns into the 4-particle amplitude for $\phi^4$ interactions, conventionally represented by the Feynman diagram
\beq
\begin{split}
\begin{tikzpicture}
\begin{feynman}
\vertex(a) {1};
\vertex[below right=of a,dot] (o){};
\vertex[above right=of o] (c){2};
\vertex[below right=of o] (b){3};
\vertex[below left=of o] (d){4};
\diagram{(a)--  (o)--  (b),(c)-- (o)--(d)};
\end{feynman}
\end{tikzpicture}
\end{split}
\eeq
Of course, for this trivial case, the Feynman diagram is evaluated as the momentum conservation $\de$-function $\de(k_1-k_2-k_3-k_4)$ multiplied by the coupling parameter. It is not surprising that a relation exists between the interaction coefficients of the classical perturbation theory in AdS, and the corresponding tree-level Feynman diagrams for a quantum field in AdS. Indeed, both objects arise at leading order in the weak coupling expansion, while the tree-level Feynman diagrams are also known to belong to the leading order of the semi-classical expansion and are thus essentially classical. For the more complicated gravitational case,
the interaction coefficients have never been recast in a form similar to (\ref{C1234}), but one may legitimately expect that such a representation exists and that it will produce structures similar to tree-level 4-graviton amplitudes in AdS. One may mention in this relation that
recent decades have seen remarkable progress in understanding
the gravitational tree level amplitudes in asymptotically Minkowski space \cite{amplitudes}.
Unfortunately, such considerations do not appear to easily generalize\footnote{I thank Yvonne Geyer for a discussion on these topics.} to the asymptotically global AdS setup that is relevant for us here.
Recent motion in that direction can be seen in \cite{adsamplitudes1,adsamplitudes2}, but focuses on AdS boundary amplitudes natural in the AdS/CFT context, rather than on the bulk amplitudes that are relevant here. Any continued progress in that direction involving bulk amplitudes would likely shed substantial further light on the patterns in 
weakly nonlinear gravitational dynamics in AdS.


\section{Turbulent dynamics}\label{sec5}

Turbulent energy transfer towards shorter wavelengths lies at the heart of the AdS instability phenomena,
as it is what makes it possible for the small energy, spread over a large volume in the small amplitude initial state, to become concentrated within a tiny area defined by its Schwarzschild radius, whereafter a small black hole forms.

Following the formulation of gravitational resonant systems in AdS in spherical symmetry \cite{FPU,CEV1,CEV2} described in section~\ref{sec3}, numerical simulations of this system in AdS$_5$ were performed in \cite{BMR}.
These simulations have confirmed the presence of a strong turbulent cascade that is a counterpart of black hole formation. Note that this approach to simulations is complementary to straightforward numerical simulations of Einstein's equations (\ref{EOMphi}-\ref{eqn:EOMConstraint}). In straightforward numerical simulations, going to very small values of $\eps$ is forbiddingly costly, as it both requires running the simulations for very long times ($1/\eps^2$) and results in formation of a very small black hole that requires a large spatial resolution. By contrast, the resonant approximation captures the universal dynamics at small $\eps$ (and $\eps$ is completely scaled out from the resonant system, so that only one simulation is required to capture the dynamics approximately for an entire range of sufficiently small values of $\eps$).
In the overlap domain where both approaches are viable, they are in excellent agreement \cite{BMR}.

What has been done in practice in \cite{BMR} is numerical simulations of the resonant system (\ref{eqn:RG2}) corresponding to AdS$_5$ starting with {\it two-mode initial data} where all $\al_n$ with $n\ge 2$ vanish.
Such two-mode initial data is a very common choice for simulations, and we shall return to it in section~\ref{sec7}. (They are in a sense the simplest initial data that lead to nontrivial evolution, since single-mode initial data remain stationary.)
While initially the energy is entirely contained in the two lowest long-wavelength modes, (\ref{eqn:RG2}) initiates energy transfer between the modes and hence the energy leaks out to shorter wavelengths.
While this phenomenon is completely generic, what is remarkable is that, for the specific values
of the mode couplings in (\ref{eqn:RG2}), this turbulent cascade takes on dramatic proportions.
As explained in \cite{BMR}, the short-wavelength part of the spectrum ($n\gg1$) can be accurately fitted (as a function of $n$) to the following expression
\begin{equation}\label{turbtail}
\al_n(\tau)\sim n^{-\sigma(\tau)} e^{-\rho(\tau)n}.
\end{equation}
Furthermore, $\rho$ vanishes at a certain finite moment in time, and $\sigma=2$ at that moment.
One thus reaches a power-law spectrum in finite time starting with two-mode initial data, and the energy
becomes essentially completely delocalized in the mode space. One may refer to this process as a {\it finite-time turbulent blowup}.

The moment when the turbulent blowup happens, as well as the detailed shape of the evolution generated by the resonant system are in agreement \cite{BMR} with the numerical simulations of Einstein's equations.
This brings us to a subtlety that deserves some mention. While the time scales that we are considering ($1/\eps^2$)
are within the validity domain of the resonant approximation, there is one more approximation that has been employed in deriving the resonant system (\ref{eqn:RG2}), namely, approximating the full gravitational
nonlinearity in the effective wave equation for the scalar by its cubic part. This is seen in our expansion
of the solutions to (\ref{eqA}-\ref{eqde}) up to quadratic order in $\phi$, which correctly recovers the cubic part of the right-hand side of (\ref{effwave}), but omits the higher order (quintic, etc) contributions that our treatment has neglected. Since the amplitude of $\phi$ is small, one expects this approximation to be valid on time scales $1/\eps^2$ (some more detailed considerations are given in \cite{extend}). However, very close to collapse, fields may locally take large values even for small initial data, so the validity of the cubic approximation to the gravitational nonlinearity in this narrow time range is, in principle, in question. (This concern should not affect the main stages of the turbulent cascade, however, since the collapse phase is very short.) It has been observed, nonetheless, that the resonant dynamics of (\ref{eqn:RG2}) approximates the full evolution surprisingly well \cite{BMR,P} despite these concerns, and this has led to a perception that black hole formation in AdS is essentially a `perturbative' process, though no complete theoretical understanding to back up this view is available at the moment. Perturbative description has been successfully applied in \cite{BM} to the related problem of planar collapse in the Poincar\'e patch of AdS (unlike all the considerations in global AdS in this review, these solutions carry an infinite amount of total energy, and the collapse process terminates in a formation of a spatially extended black object), and to collapse in global AdS in the `thin shell' regime.

It is expected that similar turbulent blowup will happen in AdS$_{d+1}$ with $d>4$, while the value of the exponent $\sigma$ in (\ref{turbtail}) is dimension-dependent. These turbulent exponents have been discussed and measured numerically in the context of the resonant approximation for some higher-dimensional cases in \cite{Deppe}, while earlier numerical work for higher-dimensional AdS collapse in the original gravitational equations can be found in \cite{JRB}. In AdS$_4$, the situation is more subtle since the resonant dynamics exhibits powerful reverse cascades that actually drive the system close to the initial state after the initial direct cascade \cite{FPU,returns}, as we shall discuss in section~\ref{sec7}. It remains an open question whether a turbulent blowup still takes place within the resonant approximation after a long  sequence of direct and reverse cascades, while in direct numerical simulations of Einstein's equations with finite perturbations, collapse to a black hole does occur after a few such direct-reverse cascade sequences. An open dispute on this topic can be consulted in \cite{FPU,FPUcomm,FPUrepl}. The case of AdS$_3$ is special \cite{BJ} as it is known a priori that black holes cannot form for sufficiently small amplitude initial data, since there is a lower bound on black hole masses. However, numerical simulations \cite{BJ} still display a decay of the `analyticity radius'
$\rho$ in (\ref{turbtail}) that can be seen as an infinite-time turbulent blowup ($\rho$ vanishes in the infinite future), in contrast to the finite-time blowup characteristic of the higher-dimensional cases.

Turbulent cascades are usually characterized in mathematical literature in terms of growth and blowup of {\it Sobolev norms}. For (\ref{eqn:RG2}), it is convenient to define Sobolev norms as 
\beq
S_\gamma=\sum_{n=0}^\infty n^\gamma\, \om_n |\al_n|^2,
\eeq
where $\gamma$ is an arbitrarily chosen exponent. By (\ref{sphercons}), $S_0$ and $S_1$ are conserved by (\ref{eqn:RG2}), while the higher Sobolev norms are free to grow, though of course, whether they do depends on the detailed dynamics. Evidently, once a power law spectrum forms as $\rho$ vanishes in  (\ref{turbtail}), higher Sobolev norms become infinite (this blowup furthermore happens in finite time).
Growth of Sobolev norms is a subtle phenomenon that requires detailed dynamical analysis (it is indeed a significant branch of modern PDE mathematics, as in the considerations of \cite{nls,GG}).
One should be careful not to conflate the presence of such growth with the mere fact that energy generically gets transferred between the modes by the resonant dynamics as in (\ref{eqn:RG2}).
Arguments have occasionally been voiced in the literature that because there are secular terms in AdS signifying energy transfer between the modes (and correspondingly, terms in the resonant system that
represent this energy transfer channel) a turbulent cascade must take place. Such arguments are certainly not valid at the face value, since one must distinguish something that may happen from something that does happen -- though of course such heuristic considerations are valuable as a stimulus to look deeper into the problem. Whether the turbulent cascade unfolds or not (given that the necessary energy transfer channels are present) may only be decided on the basis of detailed dynamical analysis of the relevant resonant system. (Mild forms of Sobolev norm growth in resonant systems (\ref{reseq}) with random interaction coefficients $C$ have been discussed in \cite{meloturbu} with a view to access the genericity of such growth.)

All the descriptions above have referred to the Einstein-scalar-field system of section~\ref{sec3} and the resonant approximation to its dynamics. For the Klein-Gordon equations of section~\ref{sec2}, by contrast, no significant turbulent behaviors have been observed. This is intuitively related to the short-wavelength asymptotics of the interaction coefficients analyzed in \cite{BMR,UV1,UV2,UV3,UV4}. The short-wavelength growth is a power law in terms of the mode numbers in all cases, but this power is higher for gravitational interactions \cite{UV1,UV3} that involve field derivatives than for contact interactions \cite{UV2,UV4}. Still, it remains an open question whether
turbulent behaviors may occur for contact interactions in higher dimensions, and some such phenomena have been reported in \cite{wavecollapse} (some evidence for turbulent blowup also exists for nonrelativistic analogs of these systems to be treated in section~\ref{sec8} in high numbers of spatial dimensions \cite{BMP,A}; considerations in recent papers \cite{nonrel1,nonrel2} are also driven to an extent by these topics).

Heuristic estimates have been given in \cite{BMR,UV1,UV3} that connect the specific power laws appearing
in the short-wavelength asymptotics of the interaction coefficients (as functions of the mode numbers) and the
power of $n$ present in (\ref{turbtail}), but there is no systematic derivation of such relations. One essential question is which specific asymptotic regimes for the four indices of $C_{nmkl}$ in (\ref{reseq}) or (\ref{reseq2}) determine turbulent behaviors. It is a natural thought \cite{P,M} that the important coefficients are not the ones with all the four mode numbers taking large values, but rather the interaction coefficients coupling some low and some high modes. Such interaction coefficients indeed facilitate energy transfer out of the long-wavelength part of the spectrum. While the situation has not been sorted out in full generality, some support for this intuitive picture comes from \cite{cascade}. In that paper, a variety of resonant systems of the form (\ref{reseq}) has been considered that display strong turbulent behaviors. In parallel, one would study the dynamics of modified resonant systems where infinite subsets of the values of $C_{nmkl}$ have been replaced by zeros, and it was observed that indeed, removing all couplings that do not involve mode 0, for example, does not weaken turbulent cascades (and sometimes even strengthens them).

Needless to say, the interaction coefficients for the simplest gravitational case defined by (\ref{Tdef}-\ref{Sdef}) are extremely complicated. In fact, in \cite{GMLL} an algorithm was developed that allows one
to evaluate all the integrals involved in (\ref{Tdef}-\ref{Sdef}) for the case of AdS$_4$ and obtain explicit functions of the mode numbers that define the interaction coefficients. These functions, however, are so
complicated that only a small part of the resulting expressions could be quoted in the published article.
From this perspective, the detailed form of the gravitational interaction coefficients is unlikely to be of much use for qualitative analytic work. Rather, the best hope is that it will be possible to relate qualitative properties of the resonant solutions (such as turbulence) to some relatively simple properties of the interaction coefficients (such as the short-wavelength asymptotics), and then prove that the actual complicated formulas for the interaction coefficients possess these simple properties. With this mindset,
it would be ideal to have a cartography of various possible behaviors of the general resonant system (\ref{reseq}) in the space of the interaction coefficients $C_{nmkl}$, in particular, in relation to turbulence.

As a modest initial step for developing this picture, we mention a few concrete simple resonant systems
given by (\ref{reseq}) that display various forms of turbulence. First, the simplest and (in some ways) most remarkable of all is the cubic Szeg\H{o} equation introduced and analyzed in \cite{GG}. This equation is defined by
\beq
C_{nmkl}=1.
\eeq
It is Lax-integrable and admits two distinct Lax pairs, as well as other powerful integrability-related structures. Integrability has been used \cite{GG} to construct its formal general solution and prove that initial conditions exist for which Sobolev norms display unbounded growth. Integrable deformations of the cubic Szeg\H{o} equation exist \cite{Xu,cascade} for which elementary explicit solutions can be constructed with exponential Sobolev norm growth, but no finite-time blowup.

In terms of finite-time blowup, a few explicit examples were proposed at the end of \cite{cascade} for which  it is known numerically that a finite-time turbulent blowup occurs starting with two-mode initial data. One of these examples is
	\beq\label{Cblowup}
	C_{nmkl} = \sqrt{(n+1)(m+1)(k+1)(l+1)}.
	\eeq
The finite-time blowup is of the same form as (\ref{turbtail}). A few further examples with similar
properties can be found in \cite{cascade}. We thus have at our disposal a few resonant systems of the form
(\ref{reseq}) with a very simple formula for the interaction coefficients that generate behaviors similar to
the very complicated gravitational resonant system (\ref{eqn:RG2}).

One further point deserves mention in relation to turbulent phenomena in resonant systems.
As emphasized in \cite{DFLY}, formation of a power-law energy spectrum $|\al_n|^2$
(or Sobolev norm blowup) is by itself insufficient to ensure that a singularity arises in the position space
profile of the corresponding fields, as should happen in a gravitational collapse.
Additionally, the phases $\varphi_n\equiv\arg\al_n$ must be locked in an approximately linear pattern
$\varphi_n(t)= A(t)n+B(t)+O(1/n)$. Such {\it phase locking} or {\it phase coherence} is indeed present in (\ref{turbtail}), and it is also commonly seen in other resonant systems \cite{A}, irrespectively of turbulence.
One way to enforce phase locking is to say that there is only one singular point at the boundary of the
convergence disk of the series $\sum_n \al_n z^n$ that defines the generating function of $\al_n$
in the complex $z$ plane. It seems likely that the emergence of phase locking in physical resonant systems
is related to the simple power-law asymptotics \cite{UV1,UV2,UV3,UV4} of the interaction coefficients at large mode numbers, though this has not been proved.


\section{Solvable dynamics}\label{sec6}

In the previous section, we have focused on violent turbulent behaviors leading to collapse, which are
difficult to get under analytic control and indeed touch some of the cutting-edge topics of contemporary PDE mathematics. To provide a counterpoint, this section reviews some situations where elementary explicit solutions of resonant systems (\ref{reseq}) can be constructed, and the simple behaviors these solutions display.

The simplest possible evolution is no evolution at all, and this is what is accomplished by stationary solutions of the form
\beq
\al_n(\tau)=e^{-i(\lambda+\mu n)\tau} A_n
\eeq
with time-independent $\lambda$, $\mu$, $A_n$. The energy spectrum $|\al_n|^2$ of such solutions is frozen in time.
Such solutions are generically present in resonant systems of the form (\ref{reseq}), as can be seen by substituting the above ansatz in that equation and observing that all the time dependencies cancel out,
leaving a system of algebraic equations for $A_n$. AdS-related stationary solutions of this form have been extensively studied in the literature \cite{FPU,UV2,UV3,GMLL,CF,BHP}. The simplest of those are single-mode solutions where only one $A_n$ is nonzero, though large families of more sophisticated solutions exist.

Solutions that do display energy transfer are more difficult to analyze. However, a number of resonant systems have been discovered\footnote{The considerations of \cite{CF,BHP} are phrased for a conformally coupled scalar in the Einstein static universe $R\times S^3$ which is closely related to the same system in AdS$_4$. The corresponding cubic conformal wave equation arises as a consistent truncation of the ${\cal N}=4$ super-Yang-Mills theory on $R\times S^3$ \cite{J}, an important model in contemporary mathematical high-energy physics.} in the course of AdS-related investigations \cite{CF,BEL,BEF,BBCE}
that manifest the following curious pattern in their evolution:
\begin{enumerate}
\item They possess exact solutions parametrized by three complex degrees of freedom $a$, $b$ and $p$ of the form
\begin{equation}
\alpha_n = f_n\, \big\{b(t) + n a(t)\big\}\, (p(t))^{n},
\label{eq:alpha_ansatz}
\end{equation} 
where $f_n$ are a sequence of time-independent numbers.
\item The dynamics of these solutions shows exactly periodic behaviors of the energy spectrum $|\al_n|^2$.
\end{enumerate}
Note that the two-mode initial data commonly used for studying turbulence are accommodated within the  ansatz (\ref{eq:alpha_ansatz}) as a nonsingular limit $p\to 0$, $ap\to\cnst$. The dynamics is in a way diametrically opposite to a turbulent blowup: instead of a violent cascade that spreads the energy over all modes, the dynamics periodically reconstructs the amplitude spectrum of the initial state with a perfect precision.

The actual resonant systems displaying the above pattern studied in \cite{CF,BEL,BBCE} originate
from non-gravitating dynamical systems in AdS and related setups. Only the case of \cite{BEF} includes gravitational interactions and results from taking a nonrelativistic limit of a gravitating scalar field in AdS$_5$. In view of the underlying algebraic structure we are about to expose, similar phenomena may be expected in relativistic gravitating systems, but only outside spherical symmetry, where analytic treatments are very challenging and have not been developed.

The algebraic structure underlying the above solvable features in resonant systems observed in \cite{CF,BEL,BBCE,BEF} has been distilled in \cite{solvable} into a huge class of resonant systems of the form (\ref{reseq}) that share the same properties. A key feature of this class of systems is that all of them
admit a conserved quantity of the form
\beq
Z  = \sum_{n=0}^{\infty}\sqrt{(n+1)(n + G)}\,\bar{\alpha}_{n+1} \alpha_n
\label{eq:Z_cq}
\eeq
in addition to the evident conserved quantities
\beq
N=\sum_{n=0}^\infty |\al_n|^2,\qquad E=\sum_{n=1}^\infty n|\al_n|^2. 
\eeq
Here, $G$ is a number assigned individually to each system in the large class of \cite{solvable}. It turns out that the conservation of $Z$ implies a finite difference equation satisfied by the interaction coefficients $C_{nmkl}$ that, in turn, guarantees that the ansatz (\ref{eq:alpha_ansatz}) is consistent provided that
\begin{equation}
f_n=\sqrt{ \frac{\left(G\right)_n}{n!}},
\label{eq:f_n_function}
\end{equation} 
where $(G)_n$ is the Pochhammer symbol. This structure captures all the cases that have emerged from AdS-related considerations, while the full class of $C_{nmkl}$ with these properties forms an infinite-parametric family. The exact periodicity of $|\al_n|^2$ then follows from restricting the dynamics to the ansatz (\ref{eq:alpha_ansatz}) and explicitly solving the resulting superintegrable Hamiltonian mechanical system \cite{solvable}. 

A natural question is under what circumstances resonant systems of the large class developed in \cite{solvable} emerge as resonant approximations to PDEs of physical interest. The answer to this question is given in \cite{breathing}. The conservation law (\ref{eq:Z_cq}) can be traced back to {\it breathing modes} of the original PDE from which (\ref{reseq}) arises as a resonant approximation. Breathing modes
are functions on the phase space whose evolution is exactly periodic with the same period for {\it all} solutions of the equations of motion. One such obvious breathing mode is the center-of-mass motion of any field system in AdS (which performs simple periodic motions independently of all other dynamical details, just like the center-of-mass in Minkowski space moves with a constant velocity). Another well-known example coming from nonrelativistic analogs of AdS systems is the Pitaevskii-Rosch breathing mode \cite{PR,qPR} for two-dimensional nonlinear Schr\"odinger equations. It may sound surprising that something as trivial as the center-of-mass motion may have nontrivial dynamical repercussions for the evolution of the other degrees of freedom. The logic, again, is that the perfect periodicity of the center-of-mass motion restricts the quartic mode couplings $C_{nmkl}$ in a way that leads to the emergence of solutions
(\ref{eq:alpha_ansatz}) as well as the other properties described above. The ansatz (\ref{eq:alpha_ansatz}) is specific to quartic nonlinearities (and thereby relies on the dynamical structure beyond the simple center-of-mass motion) and has no known generalizations to higher nonlinearities \cite{quintic}. At the same time, quartic nonlinearities are generic which explains why the structure described above arises frequently in concrete examples of practical relevance.

The perfectly periodic center-of-mass motion in AdS is responsible for the emergence of solvable structures
in the {\it maximally rotating} sector of the AdS normal mode spectrum \cite{BEL}. This sector consists of modes with the maximal projection of angular momentum on the polar axis. Such modes reside in the multiplets on the rightmost diagonal of fig.~\ref{figtower}, and they are analogous to what is known as the `lowest Landau level' in the literature on trapped Bose-Einstein condensates \cite{BBCE}. Since the center-of-mass motion will remain the same for any covariant nonlinearities, rather than just the contact interactions for a scalar field used in \cite{BEL}, it is natural to expect that the corresponding solvable structure will generalize to a large class of nonlinear wave equations. It is also likely to generalize to some gravitational setups. While in higher dimensions, gravitational waves are present and it may not necessarily be possible to find a consistent truncation of the resonant dynamics to the simple form (\ref{reseq}), in AdS$_3$ with a gravitating complex scalar field, gravity does not have any propagating degrees of freedom and can be integrated out even outside spherical symmetry, leaving an effective wave equation for the scalar matter. It is legitimate to expect that the maximally rotating sector of this theory will display solvable features of the type treated in \cite{solvable,breathing}, though no detailed analysis has been performed as yet. The setup of a complex scalar with gravitational interactions (and their higher-spin generalizations) in AdS$_3$ has received significant attention in the context of AdS/CFT correspondence \cite{hspin}. 


\section{Approximate dynamical recurrences}\label{sec7}

The previous section described situations when the evolution of the amplitude spectrum $|\al_n|^2$
due to the resonant system (\ref{reseq}) is exactly periodic, and hence perfect returns to the initial amplitude configuration occur. This situation requires very special properties of the interaction coefficients 
$C_{nmkl}$ that can be drawn from a very large family constructed in \cite{solvable}, and yet
members of these family form a tiny subset of all possible interaction coefficient choices, and are hence
highly non-generic.

It turns out, however, that the closely related phenomenon of approximate dynamical recurrences
is rather generic for systems of the form (\ref{reseq}), as will be spelled out in this section.
Approximate dynamical recurrences, namely the situations where the distribution of energy among the normal modes passes close to
the initial configuration, are a significant topic in the domain of nonlinear dynamics, where it is often
associated with the `Fermi-Pasta-Ulam paradox' (FPU) for nonlinear oscillator chains \cite{FPUrev}.
Some technical details for recurrences in resonant systems of the form (\ref{reseq}) are different from the classic nonlinear chain setup, since the recurrences occur in the fully nonlinear regime with respect to the resonant system (\ref{reseq}) that has no small parameters,
while they are seen only in the weakly nonlinear regime in the standard FPU problem. With respect
to the original PDEs from which resonant systems of the form (\ref{reseq}) arise, the recurrences are in a weakly nonlinear regime, but a highly resonant spectrum of linearized normal modes plays an essential role and indeed underlies the emergence of the resonant approximation (\ref{reseq}), while this feature is
absent in the usual FPU story. Nonetheless, both subjects broadly fit in the topic of approximate dynamical recurrences in nonlinear systems.

\begin{figure}[t]
\centering
\includegraphics[scale=0.6]{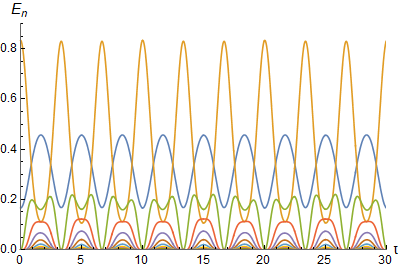}
\caption{Approximately periodic behaviors of the resonant system (\ref{eqn:RG2}) from the numerical simulations of \cite{returns}. The plot is actually not perfectly periodic, contrary to how it may appear to a naive naked eye examination. The vertical axis depicts the energies of the individual modes with the orange curve corresponding to mode 1, the blue curve to mode 0, and the remaining curves to a few higher modes.}
\label{AdS9}
\end{figure}

In relation to AdS perturbations, FPU-like phenomena have received particular attention for the gravitational resonant system (\ref{eqn:RG2}) in AdS$_4$. These FPU-like recurrences were identified in the numerical work of \cite{FPU} and later studied in more detail in \cite{returns}, typically starting with two-mode initial data. For some initial states, the initial amplitude spectrum is recovered with such precision that, looking at the plots of the amplitude spectrum evolution in a regular printout, it is impossible to distinguish the returns from perfect. An example of such slightly imperfect periodic behaviors borrowed from numerical simulations of \cite{returns} is given in fig.~\ref{AdS9}.

The approximate returns are expected to be rather generic for two-mode initial data if the turbulent cascades are not strong. They are further enhanced by the presence of rational numbers among the
interaction coefficients, which can be traced back to the relation of the AdS mode functions to simple orthogonal polynomials. Following \cite{returns}, one can give the following quick sketch that accounts for these behaviors. While far from rigorous, this sketch in fact captures the relevant dynamics, and even allows quantitative predictions in terms of which numbers of direct-reverse cascade sequences result in particularly accurate returns to the initial configuration, as demonstrated in \cite{returns} for the resonant system (\ref{eqn:RG2}) in AdS$_4$.

Imagine that, within the two-mode initial data $\al_{n\ge2}=0$, mode 0 dominates and mode 1 has a small amplitude. In this case, one expects a very weak turbulent cascade, in close relation to single-mode solutions  being stationary. It is then natural to assume a spectrum of the form
	\begin{equation}\label{SLL0defeq}
	\alpha_{n} = \delta^{n}\, q_{n}(t), 
	\end{equation}
with a small free parameter $\de$, expand the equations up to leading order in $\de$, and explore the consequences. Retaining only leading terms in the small $\de$ expansion in (\ref{reseq}) leads to
	\begin{equation}
	i\frac{d{q}_{n}}{d\tau} = \bar{q}_{0}(t)\sum_{k=0}^{n}C_{n0k,n-k}q_{k}q_{n-k}.
	\label{eq:SLL}
	\end{equation}
This equation is considerably simpler than the full system (\ref{reseq}), as it can be solved recursively mode-by-mode. We can set $q_0(0)$ and $q_1(0)$ to 1 by the scaling symmetry of (\ref{reseq}) and by a choice of the definition of $\de$, respectively. Then, under the evolution (\ref{eq:SLL}), the first two modes simply oscillate as 
\beq
q_0(t)=e^{-iC_{0000}\tau},\qquad q_1(t)=e^{-iC_{0101}\tau}.
\eeq
The higher modes $q_n$ are oscillators with eigenfrequencies $2C_{0n0n}$ and driving terms made of the lower modes. Then, recursively, solutions for $q_n$ are linear combinations of oscillatory terms $e^{-i\Omega\tau}$ with $\Omega$ given by a linear combination of $C_{0m0m}$ ($m\le n$) with integer coefficients.
It turns out that $C_{0n0n}$ are all rational in AdS$_4$ \cite{returns}. As a result, any finite set of $q$'s is guaranteed to have a common period. A basic return to the initial two-mode configuration occurs whenever $q_2(\tau)=0$, in which case the missing energy is in modes 3 and higher, and hence is of order $\de^6$.
If modes $q_2$ and $q_3$ vanish simultaneously, which is guaranteed to happen at some point in view of the common periods mentioned above, and the fact that $q_2(0)=q_3(0)=0$, a much more precise return occurs, where the missing energy is entirely in modes 4 and higher, and is hence of order $\de^8$. The returns can furthermore be made as precise as one wishes in this picture, since one can demand any range of $q$'s to vanish simultaneously, and that is guaranteed to happen at some point in the evolution. Even if there are no exact rational relations between the various
turbulent cascade frequencies, approximate common periods may exist and they will be seen as
dynamical returns of enhanced precision. One can in fact predict using such arguments which specific numbers of direct-reverse cascades reconstruct the initial state particularly accurately, and this prediction continues to hold even for comparable initial mode energies of the two-mode initial state, away from the $\de\to 0$ limit. Similar considerations can be developed for two-mode initial data dominated by mode 1 \cite{returns}.

Originally seen in numerical simulations of AdS$_4$, these FPU-like behaviors are thus expected to manifest themselves in a wide range of resonant systems of the form (\ref{reseq}). They are not seen, however,
in the dynamics of resonant systems of higher-dimensional AdS spaces, where already the first direct energy cascade leads to a turbulent blowup \cite{BMR} which terminates the resonant evolution, and the system never gets a chance to develop a reverse cascade that could lead it to the vicinity of the initial amplitude configuration.


\section{Nonrelativistic limits}\label{sec8}

Similarities between resonant approximations to AdS dynamics and analogous treatments of nonlinear Schr\"odinger equations in harmonic potentials
were pointed out in \cite{BMP}. Indeed, the tower of energy eigenstates of a quantum harmonic oscillator with an isotropic potential is exactly
identical to the AdS frequency tower given in fig.~\ref{figtower}. Consequently, the structure of the resonant approximation to such equations
is identical to the AdS case and will produce an equation of the form (\ref{reswave}), though evidently the specific values of the interaction coefficients $C$ will differ from their AdS values. The relation between the two systems has been made more explicit by deriving \cite{BEL} the nonlinear Schr\"odinger equation as a nonrelativistic limit of the AdS Klein-Gordon equation (\ref{AdSwave}). This derivation will be reviewed below.

The relation between AdS systems and their nonrelativistic counterparts is attractive from a few different angles:
\begin{enumerate}
\item It recasts the AdS dynamics and nonlinear Schr\"odinger equations into the same mathematical pattern, and thus creates a bridge between AdS studies arising from general relativity and high energy physics, and nonlinear Schr\"odinger equations in harmonic potentials. The latter are a common mathematical tool for studies of trapped ultracold atomic gases \cite{BDZ}, where the name `Gross-Pitaevskii equation' is more commonly used.
\item The PDEs involved on the nonrelativistic side are much more compact and user-friendly (though their dynamics may still be complicated).
\item Much more substantial mathematical work exists for nonlinear Schr\"odinger equations than for AdS problems. Some examples dealing directly 
with the resonant dynamics in this setting can be found in \cite{GHT,GT,GGT,fennell}. In particular, rigorous proofs of the validity of the resonant approximation \cite{GHT,fennell} exist for the nonrelativistic cases.
\end{enumerate}

Nonlinear Klein-Gordon equation in AdS reduces in the nonrelativistic regime to a nonlinear Schr\"odinger equation with a harmonic potential in the same way as
how a nonlinear Schr\"odinger equation in empty space arises from nonlinear Klein-Gordon equations in Minkowski spacetime.
Physically, this is just a transition from classical field theories describing relativistic bosons to nonrelativistic bosons.
To see the connection explicitly, we start with (\ref{AdSwave}) in AdS, now with a complex scalar field $\phi$ and the corresponding nonlinearity $|\phi|^2\phi$ on the right-hand side, and introduce $\Psi(t,r,\Omega)$ that will become the nonrelativistic wavefunction \cite{BEL}
after we take the limit:
\beq
\phi(t,x,\Omega)=\sqrt{2m}\,e^{-imt}\, \Psi(t, x\sqrt{m},\Omega).
\eeq
Taking the limit $m\to\infty$ recovers a nonlinear Schr\"odinger equation for $\Psi$ at any fixed value of $r$ with $x\equiv r/\sqrt{m}$:
\beq\label{GPHO}
i\frac{\del\Psi}{\del t}=\frac12\left(-\nabla^2+r^2\right)\Psi +|\Psi|^2\Psi.
\eeq
(A nonrelativistic limit in AdS zooms on a small neighborhood of the origin, since all small velocities in AdS result in motions of a small amplitude, in contrast to Minkowski spacetime.)

One could follow similar steps starting from a gravitating scalar in AdS$_{d+1}$ so as to end up with a Hartree equation in a harmonic potential \cite{BEF}:
\beq\label{SNH}
i\frac{\del\Psi}{\del t}=\frac12\left(-\nabla^2+|x|^2\right)\Psi -\left(\int dy\frac{|\Psi|^2}{|x-y|^{d-2}}\right)\Psi,
\eeq
in other words, an equation in which the wave function density self-gravitates with a Newtonian potential, in addition to an external harmonic potential.
Such equations have appeared in the literature under many names in different domains of research. In a gravitational context, it is natural to refer to them as Schr\"odinger-Newton-Hooke equations \cite{BEF}. Note that after the nonrelativistic limit has been taken, one can straightforwardly pursue analysis of (\ref{SNH}) outside spherical symmetry, which would have been forbiddingly complicated in the original general-relativistic context. (A similar nonrelativistic limit was considered in \cite{dSnonrel} for de Sitter, rather than AdS, spacetimes, resulting in a similar system with an inverted harmonic potential.)

A few works have pursued analysis of resonant systems emerging from nonlinear Schr\"odin\-ger equations in harmonic potentials, in relation to possible turbulent phenomena \cite{BMP}, in relation to constructing explicit solutions \cite{BEF,BBCE} that follow the general pattern of \cite{solvable}, and in relation to properties of vortices (wavefunction zeros) in stationary configurations \cite{GGT}. Rather than reviewing these publications in detail, we would like to use the simplicity of nonlinear Schr\"odinger equations to extract some pedagogical benefits and give a very short derivation of the resonant dynamics that serves as a prototype for the much more convoluted relativistic treatment of section~\ref{sec2}.

Consider a one-dimensional nonrelativistic nonlinear Schr\"odinger equation in a harmonic potential:
\begin{equation}
i\,\frac{\partial \Psi}{\partial t}=\frac12\left(-\frac{\partial^2}{\partial x^2}+x^2\right)\Psi +|\Psi|^2\Psi.
\label{NLS1d}
\end{equation}
If the nonlinearity is neglected, one is left with the linear Schr\"odinger equation of a harmonic oscillator, solved by
\begin{equation}
\Psi_{\mbox{\tiny linear}}=\sum_{n=0}^\infty \alpha_n \psi_n(x) e^{-iE_n t},\qquad E_n=n+\frac12,\qquad \frac12\left(-\frac{\partial^2}{\partial x^2}+x^2\right)\psi_n=E_n\psi_n,
\label{NLS1dlin}
\end{equation}
with constant $\alpha_n$. We then design a change of variables for the full nonlinear system based on this linearized solution:
\beq
\Psi(x,t)=\eps\sum_{n=0}^\infty \alpha_n(t) \psi_n(x) e^{-iE_n t}.
\eeq
Expressing the evolution in terms of $\al_n$ yields
\begin{equation} 
 i\,\frac{d\alpha_n }{dt}=\eps^2\sum_{k,l,m=0}^\infty C_{nmkl} \,\bar \alpha_m  \alpha_k  \alpha_l \,e^{i(E_n+E_m-E_k-E_l)t}, 
\label{NLSpreres}
\end{equation}
with $C_{nmkl}=\int dx \,\psi_n  \psi_m \psi_k \psi_l$. As usual in the time-averaging context, $\al_n$ vary slowly when $\eps$ is small, while most terms on the right hand side oscillate with periods of order 1 and hence `average out.' Discarding all such oscillatory terms leaves only terms satisfying $n+m=k+l$, whereafter switching to the slow time $\tau=\eps^2 t$ recovers our prototype resonant system (\ref{reseq}). Rigorous analysis of the validity of the resonant approximation for this specific system has been successfully undertaken in \cite{fennell}.


\section{Quantum resonant systems}\label{sec9}

This last review section takes a step aside from the previous presentation, and may be comfortably skipped by readers only interested in classical dynamics. At the same time, it supplies a powerful complementary perspective on the dynamics of resonant systems. Furthermore, this type of perspective would be unlikely to emerge from conventional systematic PDE studies.

Resonant system (\ref{reseq}-\ref{Hres}) is Hamiltonian, and as any Hamiltonian system it can be quantized following the canonical quantization procedure. This procedure replaces each complex amplitude $\al_n$ by an operator $a_n$ so that the quantum Hamiltonian corresponding to (\ref{Hres}) is
\beq
\hat H=\frac{1}2\sum_{n,m,k,l=0, \atop n+m=k+l}^\infty \hspace{-3mm} C_{nmkl} \,\ad_n\ad_m a_k a_l,
\label{ressyst}
\eeq
with daggers denoting Hermitian conjugation.
The creation-annihilation operators $\ad_n$ and $a_n$ satisfy the standard commutation relations
\beq
[a_n,\ad_m]=\de_{nm},\qquad [a_n,a_m]=[\ad_n,\ad_m]=0.
\label{comm}
\eeq
The operators act in a linear space spanned by the Fock basis to be defined below.
The main question of quantum theory is finding the eigenvectors and eigenvalues of $\hat H$. By the standard correspondence between quantum and classical dynamics, all the properties of classical trajectories may be in principle recovered from these eigenvectors and eigenvalues via the semiclassical limit. An introduction to quantum mechanics for mathematically-minded readers, and in particular the creation-annihilation operator formalism can be found in \cite{hall}.

It is customary to think that quantum dynamics is more complicated than the classical one, which is certainly true for all conventional mechanical and field-theoretic systems. This intuition is, however, misleading for quantum resonant systems (\ref{ressyst}) which are strikingly simple. Indeed, the main point here is that the diagonalization of (\ref{ressyst}) can be recast in terms of diagonalizing an infinite family of finite-sized matrices \cite{quantres}. These matrices can be diagonalized one-by-one algorithmically in a way that only requires undergraduate linear algebra. All the information on sophisticated classical behaviors of (\ref{reseq}),
turbulent, periodic, integrable, FPU-like and what-not, must be fully encoded in the results of this elementary diagonalization procedure!

The commutation relations (\ref{comm}) imply that all operators $\ad_ka_k$ commute with each other, and the eigenvalues of each of these operators are nonnegative integers, known as `particle numbers' in mode $k$. The joint eigenvectors of all of these operators
denoted $|n_0,n_1,\ldots\rangle$ form the Fock basis on which $\hat H$ acts, so that all quantum states of the system described by (\ref{ressyst}) can be constructed as linear combinations of these Fock vectors.
By definition,
\beq
\ad_ka_k|n_0,n_1,\ldots\rangle=n_k|n_0,n_1,\ldots\rangle,
\eeq
and hence, by commutation relations (\ref{comm}),
\beq
a_k|n_0,\ldots\rangle=\sqrt{n_k}|n_0,\ldots,n_k-1,\ldots\rangle,\qquad
\ad_k|n_0,\ldots\rangle=\sqrt{n_k+1}|n_0,\ldots,n_k+1,\ldots\rangle.
\label{laddact}
\eeq
The action of $\hat H$ on the Fock vectors, and hence on any quantum states, can be recovered from these relations.
The classical conservation laws of $\sum_k |\al_k|^2$ and $\sum_k k|\al_k|^2$ for (\ref{Hres}) translate in the quantum case to a statement that the operators
\beq
\hat N=\sum_{k=0}^\infty \ad_k a_k\qquad\mbox{and}\qquad \hat E=\sum_{k=1}^\infty k \, \ad_k a_k
\eeq
commute with the Hamiltonian (\ref{ressyst}). The Fock vectors $|n_0,n_1,\ldots\rangle$ are eigenvectors of these two operators:
\beq
\hat N |n_0,n_1,\ldots\rangle=\left(\sum_{k=0}^\infty n_k\right)|n_0,n_1,\ldots\rangle,\qquad \hat E |n_0,n_1,\ldots\rangle=\left(\sum_{k=1}^\infty k\,n_k\right)|n_0,n_1,\ldots\rangle.
\eeq
It is then guaranteed that the matrix elements $\langle n'_0,n'_1,\ldots|\hat H|n_0,n_1,\ldots\rangle$ vanish unless the two sets of particle numbers $\{n_k\}$ and $\{n'_k\}$ give the same values of 
\beq
N=n_0+\sum_{k=1}^\infty n_k\qquad\mbox{and}\qquad E=\sum_{k=1}^\infty k\,n_k.
\label{conspart}
\eeq
For any given $N$ and $E$, there is only a finite number of solutions of these equations for nonnegative $n_k$, which are in fact in one-to-one correspondence with integer partitions \cite{quantres}. As a result, the matrix representation of $\hat H$ in the Fock basis is block-diagonal with all the blocks being of finite sizes (these sizes may however become arbitrarily large since integer partition numbers grow without bound). Diagonalization of $\hat H$ is thus reduced to diagonalizing finite-sized numerical matrices whose entries are expressed through the interaction coefficients $C_{nmkl}$. 

We have chosen to quantize the classical resonant Hamiltonian system (\ref{Hres}) directly.
A more physically motivated alternative (though considerably more slippery mathematically)
is to quantize the original Hamiltonian PDEs that resonant systems approximate, and study them
in a weakly nonlinear regime. In this approach, taken in \cite{madagascar,shift,qperiod1,qperiod2},
the diagonalization problem for the corresponding quantum resonant system emerges from computing
the lowest order weakly nonlinear corrections to the Hamiltonian eigenvalues via a standard quantum-mechanical treatment known as the Rayleigh-Schr\"odinger perturbation theory for a degenerate spectrum.

Once the eigenvalues and eigenvectors of the resonant Hamiltonian (\ref{ressyst}) have been obtained
by straightforward matrix diagonalization, one may choose to study them further as one wishes.
One attractive perspective is provided by the {\it quantum chaos theory} \cite{haake,DKPR}.
This semi-heuristic discipline based on extensive numerical work and partial analytic insights purports
that distributions of distances between the neighboring Hamiltonian eigenvalues behave very differently
between quantum systems whose classical counterparts are integrable and chaotic. (More accurately, one is talking about the so-called unfolded level spacings; an explicit technical description can be found in \cite{quantres}.) Specifically, for classically chaotic systems, the distances between neighboring Hamiltonian eigenstates are expected to be distributed as for random matrices, that is, they must follow the Wigner-Dyson distribution. This distribution is well approximated by the so-called `Wigner surmise'
\beq
\rho(s)=\frac{\pi s}2\, e^{-\pi s^2/4},
\label{wign}
\eeq
and it vanishes at zero separation, a phenomenon known as `level repulsion.' If extra conserved quantities are present, preventing classical ergodicity, the level repulsion is expected to be washed away, leaving the Poisson distribution
\beq
\rho(s)=e^{-s},
\label{pois}
\eeq
as for distances between points randomly thrown on a line.

\begin{figure}[t]
\centering
\includegraphics[scale=0.38]{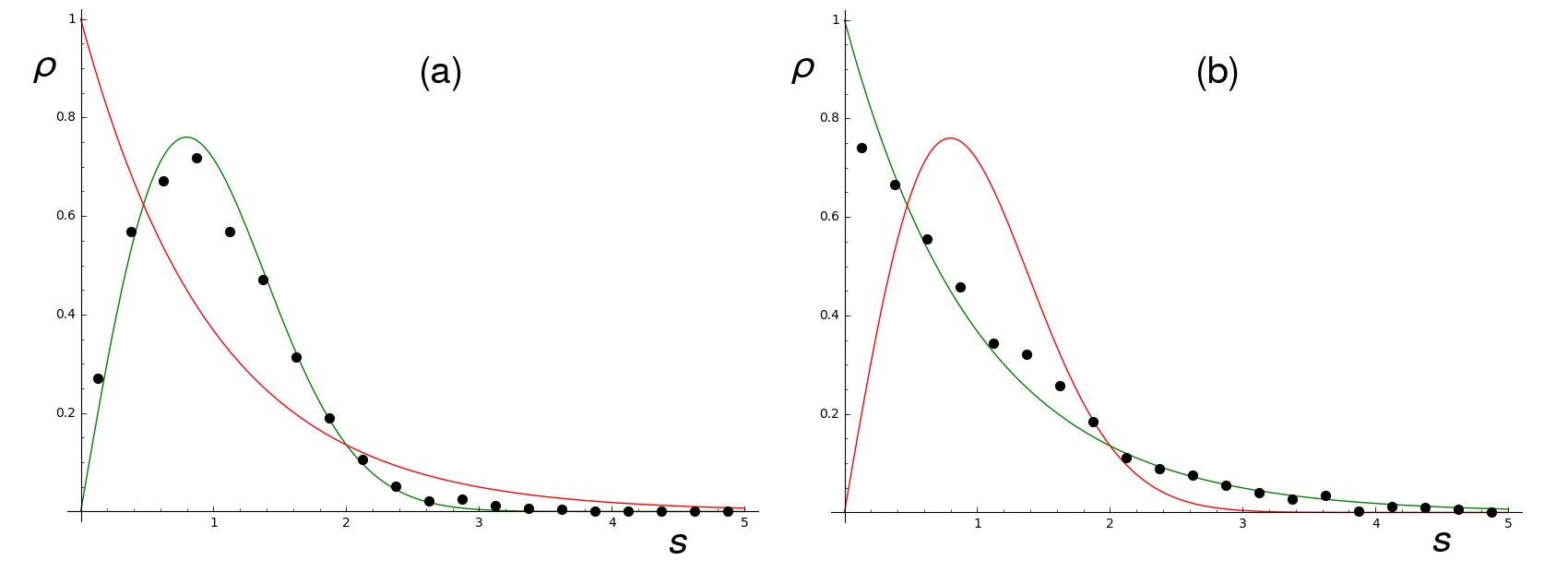}
\caption{Normalized distributions of unfolded distances $s$ between neighboring resonant Hamiltonian eigenvalues for the rotationally invariant sector of the Schr\"odinger-Newton-Hooke equation (\ref{SNH}) in (a) 3 and (b) 4 spatial dimensions at $N=E=27$, given as black dots. The bell-shaped curve is the Wigner-Dyson distribution for distances between neighboring eigenvalues of random matrices, accurately approximated by the `Wigner surmise' (\ref{wign}). The monotonically decreasing curve is the Poisson distribution (\ref{pois}).}
\label{SNH34}
\end{figure}

The lore of quantum chaos theory is beautifully validated by quantum resonant systems, as analyzed in a few distinct examples in \cite{quantres,BEF,shift}. Conversely, given a previously unexplored classical resonant system, it could be wise to run the quantum chaos diagnostic on its quantum counterpart.
Observing the Poisson distribution for the level spacings would then provide an alert that extra structures in the form of conservation laws must be present. An example of this treatment is given in fig.~\ref{SNH34} borrowed from \cite{BEF}.
This figure contrasts the distributions of unfolded distances between neighboring resonant Hamiltonian eigenvalues for the rotationally invariant sector of the Schr\"odinger-Newton-Hooke equation (\ref{SNH}) in 3 and 4 spatial dimensions in a particular $(N,E)$-block. The distribution in 3 dimensions follows the Wigner-Dyson curve and suggests that the system has no 
conserved quantities besides $N$ and $E$. The distribution in 4 dimensions is visibly closer to the Poisson curve. Indeed, an extra conserved quantity of the form (\ref{eq:Z_cq}) exists in this case.

It is natural to expect, though of course by no means guaranteed upfront, that quantum systems corresponding to the classical resonant systems with special analytic solutions in the class of \cite{solvable} treated in section~\ref{sec6} would also display pronounced analytic features. It turns out \cite{qperiod1,qperiod2} that the algebraic structures are even richer in the quantum case than what we have seen classically, and it is possible
to generate an explicit infinite family of Hamiltonian eigenstates, making the quantum problem partially solvable. It is furthermore possible to construct linear combinations within this family of Hamiltonian eigenstates that explicitly connect to the classical solutions (\ref{eq:alpha_ansatz}) via the expectation values $\langle \Psi|a_n|\Psi\rangle$, where $|\Psi\rangle$ denotes one of such linear combinations.

We finally point out that quantum resonant Hamiltonians may serve as a useful heuristic tool for guessing inequalities satisfied by the classical resonant Hamiltonians, which can be later incorporated into rigorous  mathematical work. Such bounds are commonly employed for constraining the effects of nonlinearities.
For example, the resonant system of the two-dimensional nonlinear Schr\"odinger equation in a harmonic trap (\ref{GPHO}) has a resonant Hamiltonian that satisfies, in the notation of this review, $H^2\le N^2/2$, see \cite{GHT,shift}. This bound is, in fact, equivalent \cite{GHT} to a specific inequality for Strichartz norms \cite{strnorm1,strnorm2}. A sharp version of this inequality, with a specific numerical coefficient on the right-hand side, took decades to construct \cite{GHT,strnorm3,strnorm4} by methods of functional analysis (some further related beautiful inequalities of this sort can be found in \cite{strnorm5}).
At the same time, our `quantum heuristics,' namely looking at the quantum resonant Hamiltonian and observing the patterns in its eigenvalues, would immediately suggest the correct coefficient in the sharp inequality, since by diagonalizing numerically a few concrete Hamiltonian blocks, one would notice that
the eigenvalues always fall into a specific, simply defined range. Similar ideas have been used in \cite{qperiod1,qperiod2} to construct novel lower bounds on resonant Hamiltonians, also valid in the classical theory, which are easily guessed from analyzing the patterns in quantum eigenvalues, but would be very difficult to deduce from straightforward, purely classical reasoning.


\section{Outlook}

Following the formulation of the AdS instability conjecture in \cite{BR}, considerable research into the weakly nonlinear dynamics of classical fields in AdS, with and without gravitational interactions, has been undertaken. A substantial part of this research focused on the analysis of the resonant approximation to the AdS dynamics leading to resonant Hamiltonian systems, schematically of the form (\ref{reseq}-\ref{Hres}), though analogous systems with bigger multi-dimensional sets of modes are also of relevance. These  systems were introduced in relation to AdS instability in \cite{FPU,CEV1,CEV2}. The ensuing investigations have revealed a variety of dynamical patterns, from violent turbulent cascades leading to
formation of power-law spectra in finite time, starting with predominantly long-wavelength initial data, to exactly periodic energy flows, to approximate dynamical recurrences in the spirit of the Fermi-Pasta-Ulam paradox. These findings are of interest in their own right from the standpoint of nonlinear dynamics, and connect to topics of contemporary relevance in PDE mathematics. Through the connection between AdS systems and nonlinear Schr\"odinger equations in harmonic potentials, cross-talk is established between these topics and the physics of trapped ultracold atomic gases, an active theoretical and experimental field where nonlinear Schr\"odinger equations of this form are a common mathematical tool.

The original problem of proving AdS instability for classical field systems remains unsolved, though numerical simulations of \cite{BMR} have confirmed that the turbulent cascade responsible for AdS
instability is clearly visible in the resonant system (\ref{eqn:RG2}). Rigorously demonstrating the presence of this cascade
would be the natural first step in looking for a proof of instability. This would likely require a systematic understanding of classes of resonant systems in which such turbulent blowups occur. Simple examples in the forms of toy resonant systems displaying blowup behaviors, such as (\ref{Cblowup}), are likely to provide some guidance in this regard. If the turbulent blowup has been understood, a natural next step is to establish phase-locking that would convert, along the lines of \cite{DFLY,UV3}, a power-law spectrum in the mode space to a position space singularity. Finally, one would need to analyze the horizon formation more carefully, with the full gravitational nonlinearity rather than only its cubic part retained in the resonant 
approximation, and hopefully gain control over this process in some perturbative language, possibly drawing inspiration from \cite{BM}. Needless to say, this is just a vague sketch of a possible strategy to further assault the AdS instability problem, and in any case, it remains a long way to go.

Another related problem where analytic progress has been lagging thus far is construction of resonant systems for gravitational AdS perturbations outside spherical symmetry. The case of AdS$_3$ would provide a natural starting point. The potential advantages of Hopf-like coordinates for dealing with AdS systems outside spherical symmetry likewise seem to have been underexplored. The same is true of the connections between construction of time-periodic solutions in AdS and tensor eigenvalue problems (where the tensors in question are assembled from nonlinear couplings of resonant mode quartets). The latter point also establishes contact with an active branch of modern mathematics normally thought to be far apart from nonlinear dynamics of PDEs.

\section*{Acknowlegments}

The author has been continuously funded by the CUniverse research promotion project (CUAASC) at Chulalongkorn university, and has also received support from FWO-Vlaanderen through project G006918N, from Vrije Universiteit Brussel through the strategic research program High-Energy Physics, 
and from Polish National Science Centre through grant number 2017/26/A/ST2/00530.


\end{document}